\documentclass[letterpaper,11pt]{article}

\usepackage{fancyhdr}
\usepackage{amsmath}
\usepackage{amsbsy}
\usepackage{amssymb}
\usepackage{amsthm}
\usepackage{amscd}
\usepackage{amsfonts}
\usepackage{supertabular}
\usepackage{graphics}
\usepackage{graphicx}
\usepackage{verbatim}
\usepackage{subfigure}
\usepackage{epsfig}
\usepackage{xspace}
\usepackage{euscript}
\usepackage{alltt}
\usepackage{boxedminipage}
\usepackage{float}
\usepackage{times}
\usepackage[colorlinks]{hyperref}
\usepackage[square,numbers]{natbib}
\usepackage{setspace}

\usepackage{algorithm}
\usepackage{algorithmic}

\begin{document}

\title{A nonlocal contact formulation for confined granular systems}
\author{Marcial Gonzalez and Alberto M. Cuiti\~{n}o \\
        Department of Mechanical and Aerospace Engineering, Rutgers University, \\
        98 Brett Road, Piscataway, NJ 08854, USA}

%\label{firstpage}

\maketitle

\begin{abstract}

We present a nonlocal formulation of contact mechanics that accounts for the interplay of deformations due to multiple contact forces acting on a single particle. The analytical formulation considers the effects of nonlocal mesoscopic deformations characteristic of confined granular systems and, therefore, removes the classical restriction of independent contacts. This is in sharp contrast to traditional contact mechanics theories, which are strictly local and assume that contacts are independent regardless the confinement of the particles. For definiteness, we restrict attention to elastic spheres in the absence of gravitational forces, adhesion or friction. Hence, a notable feature of the nonlocal formulation is that, when nonlocal effects are neglected, it reduces to Hertz theory. Furthermore, we show that, under the preceding assumptions and up to moderate macroscopic deformations, the predictions of the nonlocal contact formulation are in remarkable agreement with detailed finite-element simulations and experimental observations, and in large disagreement with Hertz theory predictions---supporting that the assumption of independent contacts only holds for small deformations. The discrepancy between the extended theory presented in this work and Hertz theory is borne out by studying periodic homogeneous systems and disordered heterogeneous systems.

\end{abstract}

\section{Introduction}

The understanding of confined granular systems is crucial for many fields of science and engineering. Compaction of granular media plays a relevant role in the design, optimization and control of many pharmaceutical, detergent, food, ceramic and metallurgical manufacturing processes, so much so that mechano-chemical properties of powder compacts have direct impact on the end-product performance (see, e.g., Alderborn and Nystrom \cite{Alderborn-1996}, and references therein). In the context of earth sciences, the prediction of macroscopic behavior from microscopic details of densely packed systems is a central focus of soil mechanics (see, e.g., Wood \cite{Wood-1990}).

Granular systems, such as powders or soils, are made of discrete particles larger than 1-10~$\mu$m that interact with each other through contact forces. Despite their apparent simplicity, the theoretical and computational modeling of granular media remains an active area of research to this day. Since the macroscopic behavior of these systems is fundamentally encoded at the granular scale, one of the challenges is to develop predictive and computationally efficient macroscopic models based solely on the bulk mechanical properties and the geometric configuration of the individual particles.

The static macroscopic behavior of confined granular systems (e.g., densification and strengthening) is usually modeled using either continuum mechanics or particle mechanics approaches. Models based on a continuum mechanics description of the granular system rely on phenomenological constitutive models such as Cam-Clay \cite{Borja-1990} and Drucker-Prager \cite{Drucker-1952} which assume that the strength of the compact is only a function of its porosity, or on micromechanical constitutive models derived by homogenization techniques (see, e.g., Cambou \emph{et al.} \cite{Cambou-2009} and references therein). The former are computationally efficient though only predictive after case-by-case experimental calibrations; the latter are mathematically challenging due to the heterogeneous nature of deformation at the granular scale. Models based on a particle mechanics approach, generally referred to as discrete element methods (DEM) \cite{Cundall-1979}, improve the description at the particle level but their predictability relies on the contact formulations employed. In general, at small deformations and low relative densities, Hertz theory is used to describe the contact behavior between two elastic particles \cite{Timoshenko-1970}, and elasto-plastic contact laws have been proposed to account for plastic deformations (see, e.g., Vu-Qouc and Zhang \cite{Vu-Quoc-1999}). It bears emphasis that contact formulations are also relevant to all micromechanical constitutive models. Recently, a multiscale model that seamlessly bridges the particle mechanics approach and the continuum mechanics approach has been proposed \cite{Koynov-2011,Zheng-2002}. This quasi-discrete model combines a detailed description of the granular scale with the computational efficiency typical of finite-element discretizations of continuum models.

DEM remains an attractive approach because of its ability to describe the complex kinematics of confined granular systems (e.g., particles rearrangement). However, the need of predictive contact formulations at large deformations and high relative densities has led to new approaches where contact of individual particles is described with finite-element models. Then, either all particles in the system are simulated \cite{Gethin-2003,Procopio-2005}, or force-displacement contact laws are extracted for DEM calculations \cite{Harthong-2009}. While both methodologies provide an accurate description of particles' deformations, the latter requires a case-by-case characterization of pair interactions, and the former restricts the analysis to a small number of particles.

It is interesting to note that, while the deformation of elastic spheres under contact stresses has been extensively studied, no particular attention has been devoted to the interplay of deformations due to multiple contact forces acting on a single particle. The general assumption made is that contacts between particles are independent and therefore contact forces are resolved locally. However, the deformation of an elastic sphere under contact load is not strictly local. This was early recognized by Tatara \cite{Tatara-1989} who studied two elastic spheres under compression and proposed analytical formulae that account for the effect of the reactions on the mutual surface of contact. In the context of confined granular systems, Mesarovic and Fleck \cite{Mesarovic-2000} studied elasto-plastic spheres and concluded that the assumption of independent contacts only holds in the early stages of compaction. Recently, Harthong \emph{et al.} \cite{Harthong-2009} fitted, from finite-element simulations of simple periodic lattices, contact laws that account for nonlocal-deformation effects during high-density compaction of elasto-plastic particles.

The work presented in this paper is concerned with the development of a nonlocal formulation of contact mechanics that accounts for the interplay of deformations due to multiple contact forces acting on a single particle. The analytical formulation considers the effects of nonlocal mesoscopic deformations characteristic of confined granular systems and, therefore, removes the classical restriction of independent contacts. This is in the spirit of \cite{Tatara-1989} and is in sharp contrast to traditional contact mechanics theories, which are strictly local and assume that contacts are independent regardless the confinement of the particles. For definiteness, we restrict attention to elastic spheres in the absence of gravitational forces, adhesion or friction. Hence, a notable feature of the nonlocal formulation is that, when nonlocal effects are neglected, it reduces to Hertz theory. Furthermore, we show that, under the preceding assumptions and up to moderate macroscopic deformations, the predictions of the nonlocal contact formulation are in remarkable agreement with detailed finite-element simulations and experimental observations, and in large disagreement with Hertz theory---supporting that the assumption of independent contacts only holds for small deformations.

The paper is organized as follows. The analytical nonlocal contact formulation is presented in Section~\ref{Section-Formulation} and is validated, with experimental observations and detailed finite-element simulations, in Section~\ref{Section-Validation}. In Section~\ref{Section-MacroBehavior} we investigate the discrepancy between Hertz theory and the extended theory developed in this work. Specifically, we study different dimensional configurations, different packings and different volumetric confinements of periodic homogeneous systems, and the consolidation and compaction of confined disordered heterogeneous systems. Finally, a summary and concluding remarks are collected in Section~\ref{Section-Summary}.

\section{Nonlocal mesoscopic deformations in compressed elastic spheres}
\label{Section-Formulation}

\begin{figure}[htb]
    \centering
    \includegraphics[scale=0.54]{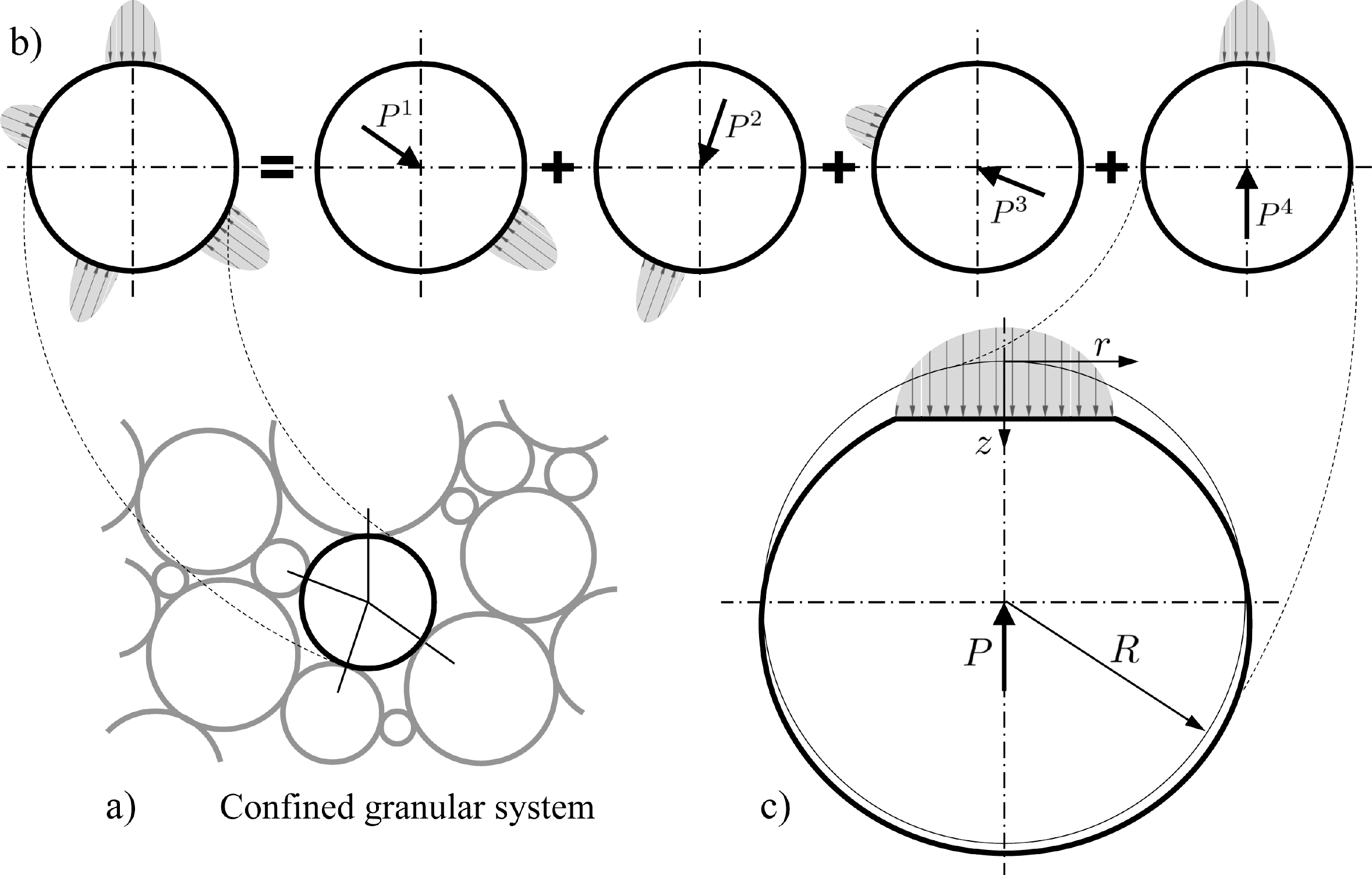}
    \caption{a) Particle within a confined granular system under general loading-conditions. b) Decomposition, by invoking the principle of superposition, of an elastostatic problem into four single-loaded boundary-value problems. c) Schematic of the deformed configuration of an elastic sphere under the action of contact stresses and supported at its center---albeit under large deformations for illustrative purposes.}
    \label{Fig-Superposition}
\end{figure}

The classical solution to the static contact problem of elastic spheres is given by Hertz theory. In the absence of gravitational forces, adhesion or friction, the theory considers linear-elastic isotropic smooth spherical particles that deform, under normal contact forces, to accommodate a flat contact surface \cite{Johnson-1985}. Additionally, the size of the contact surface is assumed to be much smaller than the dimensions and radii of curvature of the contacting bodies. Hertz theory thus applies to contact deformations that remain localized to a small area and that do not perturb other contact forces acting on the same particle. In the context of confined granular systems, this restriction is typically referred to as independent contacts. In this work we overcome this limitation by accounting for the interplay of deformations due to multiple contact forces acting on a single particle. Specifically, we consider the contribution to the contact interface of the nonlocal mesoscopic deformations induced by all other contact forces acting on the particles. This goal is achieved by invoking the principle of superposition and solving the boundary-value problem of contact stresses acting on a small region of an otherwise traction-free elastic sphere supported at its center. Figure~\ref{Fig-Superposition} depicts a particle under general loading-conditions and the decomposition of the elastostatic problem into four single-loaded boundary-value problems. A schematic of the deformed configuration of an elastic sphere under the action of contact stresses and supported at its center is also shown in the figure---albeit under large deformations for illustrative purposes.

The exact solution of the problem depicted in Figure~\ref{Fig-Superposition}c was recently reported by Zhupanska \cite{Zhupanska-2011}. She uses a general solution for the axisymmetric problem of an elastic sphere and a dual series equation approach to reduce the original boundary-value problem to a Fredholm integral equation of the second kind, which is then solved numerically. The analytical solution is achieved without replacing the contact area of the sphere by an elastic half-space (relaxing Hertz's assumption of small contact area) and it allows for determining the contact pressure and the displacement of the traction-free surface. The results presented in \cite{Zhupanska-2011} show that the spherically-distributed contact pressure predicted by Hertz theory remains accurate for relatively large contact areas. The aim of our work, however, is to solve an approximate problem that not only allows for accounting the interplay of deformations due to multiple contact forces acting on a single particle, but also that admits an efficient closed-form solution.

Next, in Section~\ref{SubSection-SingleLoaded}, we propose an approximate problem to the single-loaded boundary-value problem that is amenable to an efficient closed-form solution, and we assess its accuracy with respect to the exact solution reported in \cite{Zhupanska-2011}. In Section~\ref{SubSection-NonlocalFormulation} we present a nonlocal contact formulation that employs this approximate solution and, therefore, by invoking the principle of superposition, it accounts for the interplay of deformations due to multiple contacts acting on a single particle. Finally, in Section~\ref{SubSection-Implementation}, we provide some details about the computational implementation of the nonlocal formulation.

\subsection{Single-loaded boundary-value problem}
\label{SubSection-SingleLoaded}

\begin{figure}[htbp]
    \centering
    \begin{tabular}{cc}
    \includegraphics[scale=0.35]{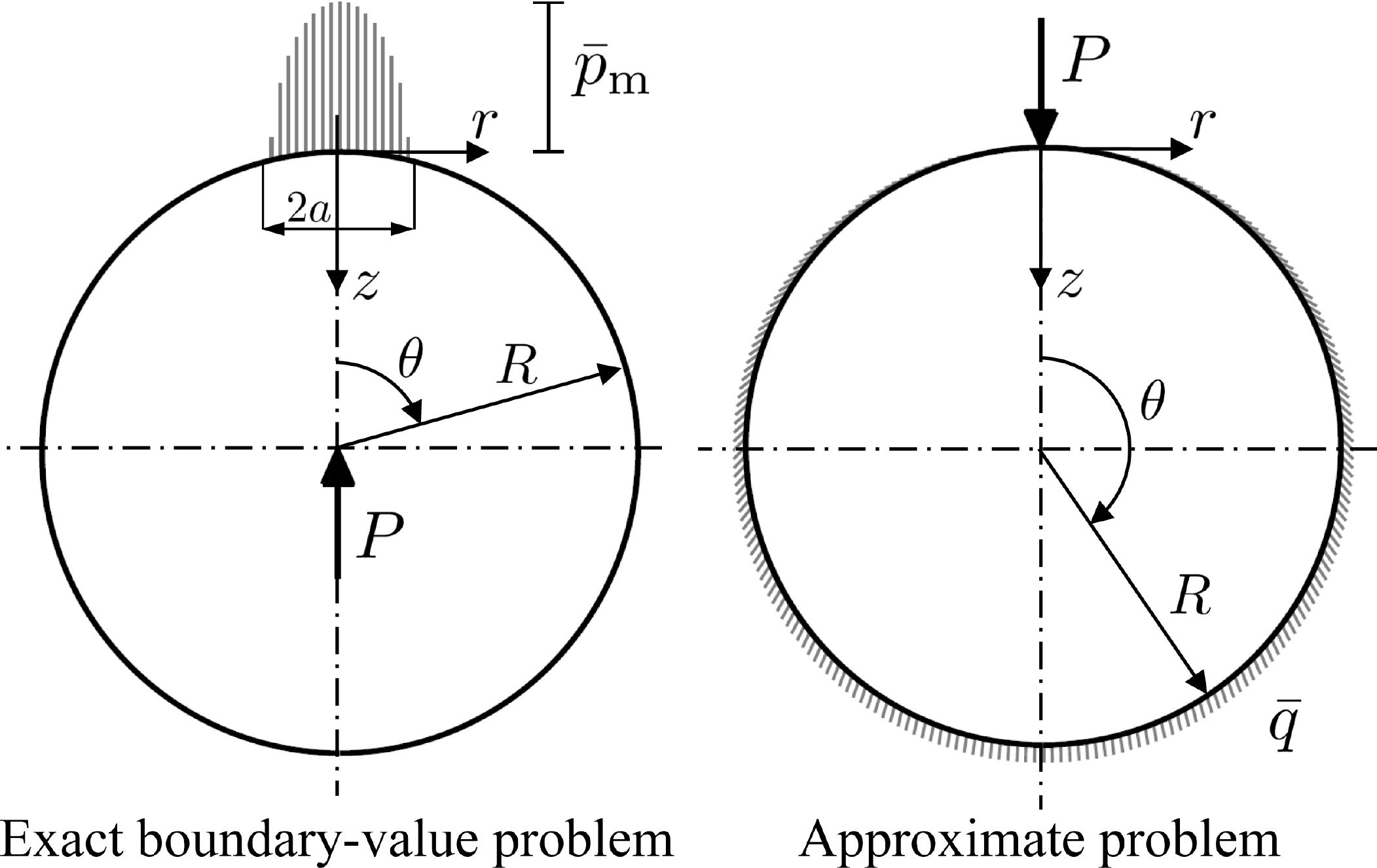}
    &
    \includegraphics[scale=0.54]{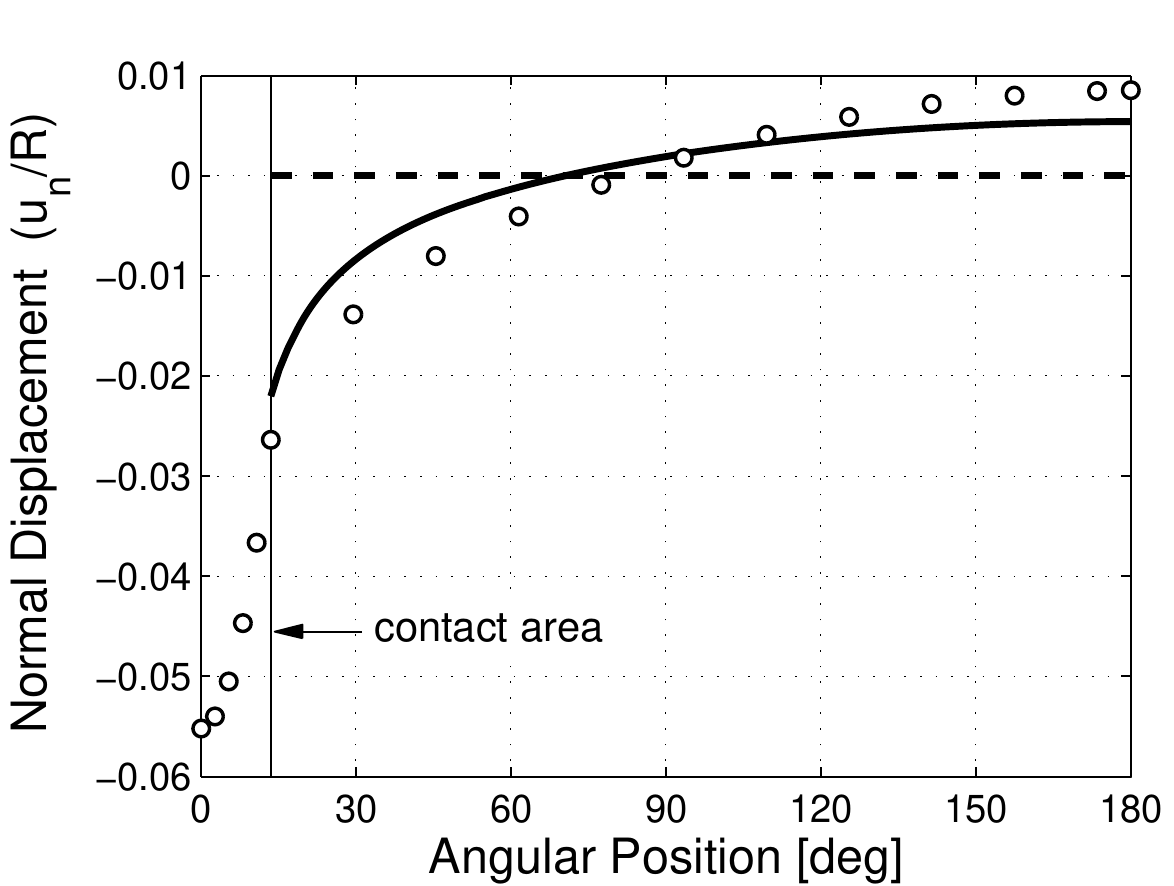}
    \end{tabular}
    \caption{Elastic sphere under the action of contact stresses and supported at its center, for $a/R=0.233$ and $\nu=0.50$. Left: Loading configurations of exact and approximate problems---the surface traction scales as $\|\bar{q}\|/\bar{p}_\mathrm{m} = \mathcal{O}(a^2/R^2)$. Right: Surface normal displacement given by the exact solution (symbols), the approximate solution (solid curve), and by traditional contact mechanics theories (dashed curve). The exact solution is taken from Zhupanska \cite{Zhupanska-2011}.}
    \label{Fig-ApproxProblem}
\end{figure}

We consider an elastic sphere of radius $R$ that deforms to accommodate a flat circular contact surface of radius $a$ under the action of a distributed contact pressure with maximum value $\bar{p}_\mathrm{m}$ and supported at its center by a concentrated force $P$. We adopt cylindrical coordinates $(z,r)$, with the $z$ axis aligned with the loading direction, and we denote a surface point by its angle $\theta$, as depicted on the left side of Figure~\ref{Fig-ApproxProblem}. The main approximations that allow for an efficient closed-form solution are then twofold:
\begin{enumerate}
  \item Displacements and tractions of the contact surface, i.e., $0 \le \theta \le \arcsin(a/R)$, are described by Hertz theory and therefore approximated by the solution of a semi-infinite elastic body under the action of a spherically-distributed contact pressure with maximum value $\bar{p}_\mathrm{m}=3P/2\pi a^2$. Thus, according to the theory of elasticity, in particular, the Boussinesq solution \cite{Johnson-1985,Timoshenko-1970}, the compressive surface displacement $u_z$ within the loaded region is given by
      \begin{equation}
        u_z(\theta) = \frac{3P(1-\nu^2)}{8 a^3 E} \left( 2 a^2 - R^2 \sin^2(\theta) \right)
      \label{Eqn-LocalDisplacement}
      \end{equation}
      where $\theta \in [0;\arcsin(a/R)]$, $E$ is the Young modulus, and $\nu$ is the Poisson's ratio.
  \item Displacements of the traction free surface, i.e., $\arcsin(a/R) < \theta \le \pi$, are approximated by the solution of a spherical particle under a concentrated force $P$---applied at the origin of coordinates and aligned with the $z$ axis---and supported by a small surface traction
      $\bar{q}(\theta) = (\bar{q}_z, \bar{q}_r)$ given by
      \begin{equation}
        \bar{q}(\theta)
        =
        \frac{P}{8\pi R^2}
        \left(
            -3 \sin(\theta/2)
            ,
            \frac{1-2\nu}{2\tan(\theta/2)(1+\sin(\theta/2))}-\frac{3\cos(\theta/2)}{2}
        \right)
      \label{Eqn-SurfaceTraction}
      \end{equation}
      Thus, the normal displacement of a surface point is
      \begin{equation}
        u_n(\theta)
        =
        \frac{1+\nu}{4\pi R E} \frac{P \left[ -2(1-\nu)-2(1-2\nu)\sin(\theta/2)+(7-8\nu)\sin^2(\theta/2) \right]}{\sin(\theta/2)}
        \label{Eqn-NonlocalDisplacement}
      \end{equation}
      where $\theta \in (\arcsin(a/R);\pi]$ and the same displacement reference frame employed above is chosen (i.e., for this case, $u_z(\pi) \rightarrow 0$ as $R \rightarrow \infty$). It is worth noting that $\bar{q}$ scales as $\|\bar{q}\|/\bar{p}_\mathrm{m} = \mathcal{O}(a^2/R^2)$, that is the surface traction is much smaller than contact forces. We present in the \nameref{Section-Appendix} a detailed derivation of this result together with other possible approximate problems amenable to closed-form solutions.
\end{enumerate}

Under the preceding assumptions, the displacement of surface points outside the contact area can be readily computed. We then proceed to compare this approximate solution with the exact solution presented by Zhupanska \cite{Zhupanska-2011}. The left side of Figure~\ref{Fig-ApproxProblem} illustrates the loading configuration of the exact boundary-value problem and the approximate problem for $a/R=0.233$ and $\nu=0.50$. It is evident that the surface traction is much smaller than contact forces even for a non-small contact area. Since an incompressible material, i.e., $\nu=0.50$, experiences the largest displacements, this is the most interesting case for assessing the accuracy of the approximation. The right side of the figure shows the surface normal displacement given by the exact and approximate solutions, and by traditional contact mechanics theories---$a^3=3PR(1-\nu^2)/4E$ is used in the representation \cite{Johnson-1985}. The results are in very good agreement and put into evidence that the interplay of deformations due to multiple contact forces acting on a single particle is not necessarily negligible.

\subsection{Nonlocal contact formulation}
\label{SubSection-NonlocalFormulation}

\begin{figure}[htbp]
    \centering
    \includegraphics[scale=0.52]{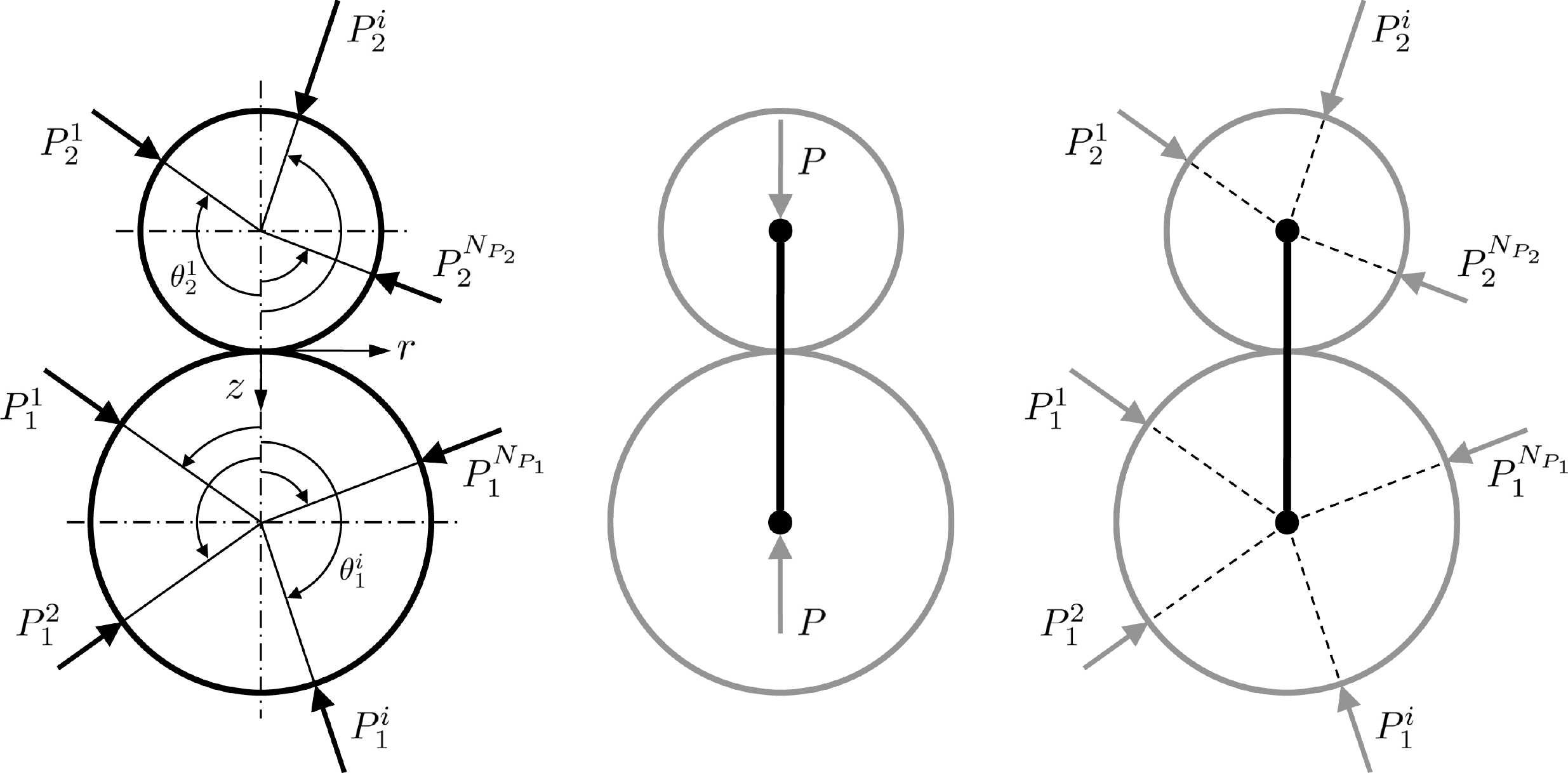}
    \caption{Two spheres in contact under a general configuration of multiple concentrated forces. Left: Schematic of the loading configuration with all forces rotated about the $z$ axis and represented on the $zr$ plane. Middle: Contact loading-conditions for Hertz theory. Right: Contact loading-conditions for the nonlocal contact formulation proposed in this work.}
    \label{Fig-TwoSpheres}
\end{figure}

We now consider the displacement of the contact surface between two elastic spheres of radius $R_1$ and $R_2$, and material properties $E_1$, $\nu_1$ and $E_2$, $\nu_2$, respectively. The spheres are compressed by a general loading configuration of concentrated forces that are statically equivalent to an effective force $P$ aligned with the $z$ axis, as depicted in Figure~\ref{Fig-TwoSpheres}. Following Hertz theory, the effective force is assumed spherically-distributed within a flat circular contact surface of radius $a$. Thus, a point within the contact surface verifies the following compatibility equation
\begin{equation}
    \gamma
    =
    R_1 - \sqrt{R_1^2-r^2}
    +
    R_2 - \sqrt{R_2^2-r^2}
    +
    w_1(r) + w_2(r)
    - \sum_{k=1,2}\sum_{i=1}^{N_{P_k}} w_{P_k^i}(r)
    \label{Eqn-Compatibility}
\end{equation}
where $\gamma$ is the relative displacement along the $z$ axis of the center of masses, and $w_1$, $w_2$ are the displacements due to the \textit{local} contact pressure. By invoking the principle of superposition, $w_{P_k^i}$ are the contributions of the \textit{nonlocal} mesoscopic-deformations induced by all other concentrated forces $P_k^i$ acting on each spherical particle $k$.

The compatibility equation \eqref{Eqn-Compatibility} then provides the equilibrium configuration $\gamma$ and the contact surface radius $a$ for a given loading configuration $P_k^i$, $\theta_k^i$. The evaluation of $w_k(r)$ is approximated by the solution of a semi-infinite elastic body under a spherically-distributed contact pressure---i.e., $w(r)=u_z(\arcsin(r/R))$ from Equation~\eqref{Eqn-LocalDisplacement}. In addition, $w_{P_k^i}(r)$ is approximated by a single value equal to the normal displacement induced along the $z$ axis by the force $P_k^i$, that is Equation~\eqref{Eqn-NonlocalDisplacement},
\begin{equation}
    w_{P_k^i}
    \simeq
    \frac{1+\nu_k}{4\pi R_k E_k}
    \frac{P_k^i \left[ -2(1-\nu_k)-2(1-2\nu_k)\sin(\theta_k^i/2)+(7-8\nu_k)\sin^2(\theta_k^i/2) \right]}{\sin(\theta_k^i/2)}
    \label{Eqn-NonlocalContribution}
\end{equation}
where $\theta_k^i$ is the angle between the concentrated force and the $z$ axis (see Figure~\ref{Fig-TwoSpheres}). Finally, the profile of the undeformed spherical cap is replaced by the first term of its series expansion
\begin{equation}
    R_1 - \sqrt{R_1^2-r^2}
    =
    R_1 \left[\frac{1}{2} \frac{r^2}{R_1^2} + \mathcal{O}\left(\frac{a^4}{R_1^4}\right) \right]
    \simeq
    \frac{1}{2 R_1}r^2
\end{equation}

After introducing the preceding approximations, equilibrium configuration for $\gamma$ and $a$, given the loading configuration statically equivalent to $P$, can be readily solved
\begin{align}
    \gamma
    &=
    \left( \frac{P}{n_H} \right)^{2/3} - \gamma_{NL}
    \label{Eqn-GammaNL}
    \\
    a^3
    &=
    \frac{3 P}{4} \left( \frac{1-\nu_1^2}{E_1} + \frac{1-\nu_2^2}{E_2} \right) \left( \frac{1}{R_1} + \frac{1}{R_2} \right)^{-1}
    \label{Eqn-aNL}
\end{align}
where $n_H$ corresponds to Hertz theory and $\gamma_{NL}$ accounts for the nonlocal loading terms (i.e., all $w_{P_k^i}$ defined by \eqref{Eqn-NonlocalContribution})
\begin{equation}
    \begin{aligned}
        \frac{1}{n_{H}}
        &=
        \frac{3}{4}
        \left( \frac{1-\nu_1^2}{E_1} + \frac{1-\nu_2^2}{E_2} \right)
        \left( \frac{1}{R_1} + \frac{1}{R_2} \right)^{1/2}
        \\
        \gamma_{NL}
        &=
        \sum_{k=1,2} \sum_{i=1}^{N_{P_k}}
        w_{P_k^i}
    \end{aligned}
    \label{Eqn-CoefficientsNL}
\end{equation}
Therefore, if nonlocal terms are neglected, the formulation reduces to Hertz theory, i.e., $P(\gamma)=n_H \gamma^{3/2}$, and the elastic contact depends on the relative position of the center of masses regardless the loading configuration (see the middle part of Figure~\ref{Fig-TwoSpheres}). In sharp contrast, if nonlocal terms are considered, the formulation departs from Hertz theory and the elastic contact now depends on the loading configuration of each sphere (see the right part of Figure~\ref{Fig-TwoSpheres}).

The formulation presented above, i.e., Equations \eqref{Eqn-GammaNL}, \eqref{Eqn-aNL} and \eqref{Eqn-CoefficientsNL}, is explicit provided the network of contact forces and loads are known. However, this is not the focus of this study. Instead, we seek for the equilibrium configuration of a system of particles given loads and static boundary conditions. That is, we aim at determining the relative displacements $\gamma$ that generate a network of forces $P$ in static equilibrium with the loads and boundary conditions. Hence, we first rewrite Equation~\eqref{Eqn-GammaNL} as follows
\begin{equation}
    P(\gamma) = n_H \left( \gamma + \gamma_{NL} \right)^{3/2}
    \label{Eqn-ForceNL}
\end{equation}
to obtain an explicit formulation in terms of relative displacements. Then, the solution of the non-linear system can be efficiently obtained by a few fixed point iterations of $P$, starting with $\gamma_{NL}=0$. It bears emphasis that $\gamma_{NL}$ is a function of the contact forces and, in the context of a numerical iterative procedure, it has to be approximated by the values of its previous numerical guess.

\begin{figure}[htbp]
    \centering
    \includegraphics[scale=0.52]{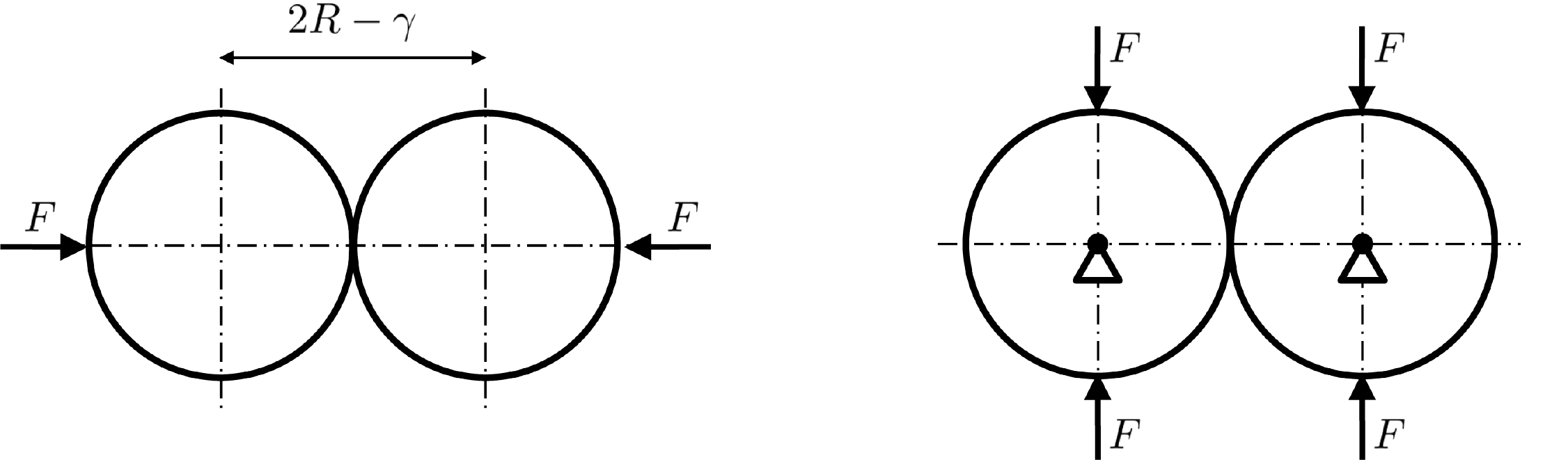}
    \caption{Two examples that illustrate nonlocal mesoscopic-deformation effects in contact pair-interactions. Left: Two elastic spheres in compression behave stiffer than Hertz theory predictions. Right: Two elastic spheres are brought into contact by transverse loading.}
    \label{Fig-Examples}
\end{figure}

It is interesting to verify that Equations~\eqref{Eqn-ForceNL} and \eqref{Eqn-CoefficientsNL} simplify to the results obtained by Tatara \cite{Tatara-1989} when two elastic spheres are compressed by forces $F$ aligned with the $z$ axis (see left side of Figure~\ref{Fig-Examples}). Then, for a loading configuration given by $P_1^1=P_2^1=F$ and $\theta_1^1=\theta_2^1=\pi$, the contribution of the nonlocal forces is
\begin{equation}
    \gamma_{NL} = \sum_{k=1,2} \frac{F(1+\nu_k)(3-2\nu_k)}{4\pi R_k E_k}
\end{equation}
and three consecutive iterations of the force $F$ are
\begin{equation}
\begin{aligned}
    F &= n_H \gamma^{3/2} + \mathcal{O}(\gamma^2)
    \\
    F &= n_H \gamma^{3/2} + \frac{3 n_H^2}{2 n_{NL}} \gamma^2 + \mathcal{O}(\gamma^{5/2})
    \\
    F &= n_H \gamma^{3/2} + \frac{3 n_H^2}{2 n_{NL}} \gamma^2 + \frac{21 n_H^3}{8 n_{NL}^2} \gamma^{5/2} + \mathcal{O}(\gamma^{3})
\end{aligned}
\end{equation}
with $n_{NL} = F/\gamma_{NL}$. Since the first iteration corresponds to $\gamma_{NL}=0$, the value of $F$ coincides with Hertz theory prediction. These are indeed the formulae obtained by Tatara in his work\footnote{Besides a (perhaps typographic) error in the third term of the series expansion.}.

It is also interesting to observe that there are loading configurations that result in compressive forces $P>0$ with no relative displacement of the particles, i.e., $\gamma=0$. This is the case of the example depicted on the right side of Figure~\ref{Fig-Examples} where the loading configuration is given by $P_1^1=P_1^2=P_2^1=P_2^2=F$ and $\theta_1^1=\theta_1^2=\theta_2^1=\theta_2^2=\pi/2$. Specifically, the reaction $P$ at the supports that enforce $\gamma=0$ is given by
\begin{equation}
    P = \frac{n_H}{n_{NL}^{3/2}} F^{3/2}
\end{equation}
where $n_{NL} = F/\gamma_{NL}$ is
\begin{equation}
    \frac{1}{n_{NL}}
    =
    \sum_{k=1,2} \frac{(1+\nu_k)(3\sqrt{2}-4-4\sqrt{2}\nu_k+8\nu_k)}{4\pi R_k E_k}
\end{equation}
It is worth noting that Hertz theory predicts $P=0$ for this configuration, regardless the value of $F$.

\subsection{Implementation of the nonlocal contact formulation}
\label{SubSection-Implementation}

\begin{figure}[htbp]
    \centering
    \includegraphics[scale=0.53]{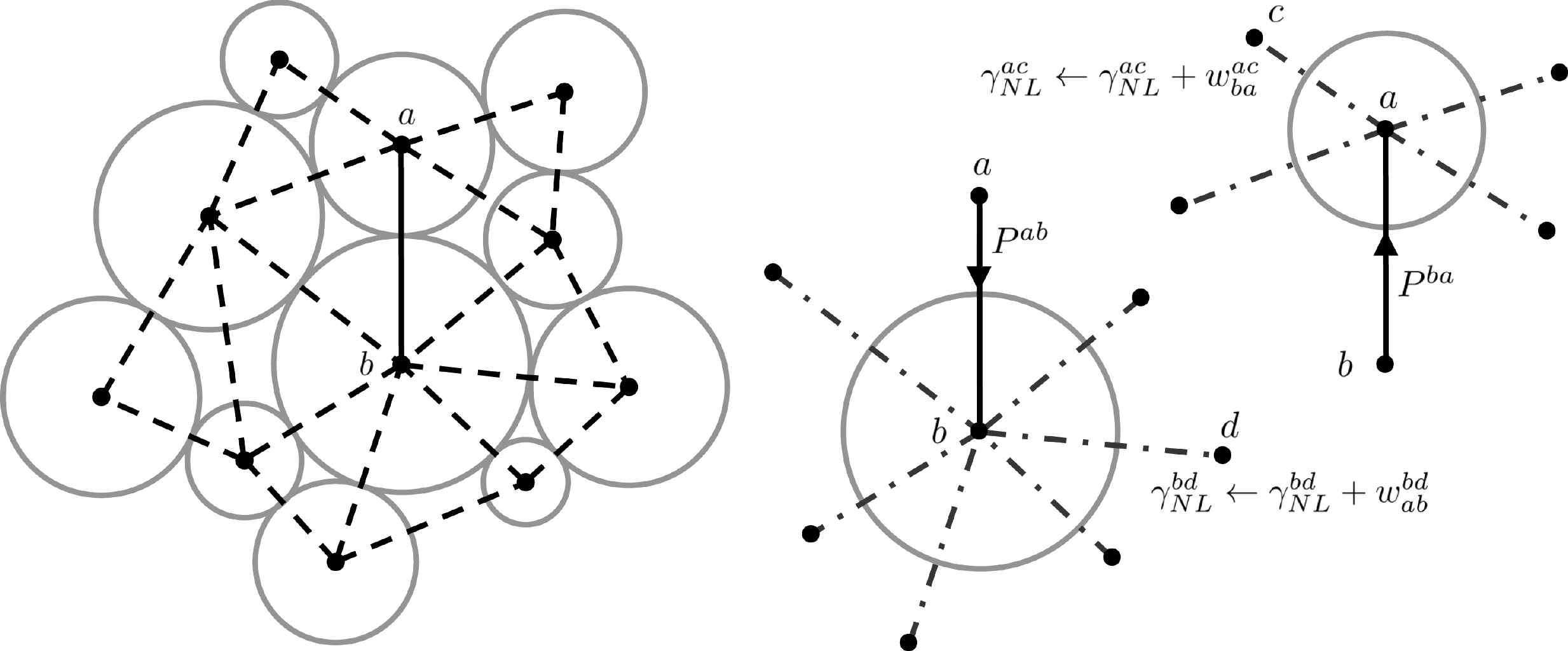}
    \caption{Left: Heterogeneous granular system and its corresponding simplicial complex. Vertices represent centers of mass and edges represent plausible contact pair-interactions (edge $ab$ is highlighted). Right: Incident edges to edge $ab$ are grouped by those connected to vertex $a$ and those connected to vertex $b$, that is those contact pair-interactions acting on particle $a$ and those acting on particle $b$.}
    \label{Fig-Algorithm-Incidence}
\end{figure}

The implementation of the nonlocal contact formulation presented in this section requires the solution of a set of non-linear equations. The set of non-linear equations for a heterogeneous granular system is uniquely defined by the material properties and geometry of the individual particles, a simplicial complex $X$ comprised by a set of vertices $E_0(X)$ that represent the center of mass of each particle and by a set of edges $E_1(X)$ that represent plausible contact pair-interactions, and the coordinates $q_{e_\alpha}$ of all vertices $e_\alpha \in E_0(X)$. The left side of Figure~\ref{Fig-Algorithm-Incidence} depicts a simplicial complex and the underlying heterogeneous granular system; additionally, vertices $a$ and $b$, and edge $ab$ are highlighted.

The solution of the non-linear system can be efficiently obtained by a few fixed point iterations of the nonlocal terms $\{\gamma_{NL}^{e_\alpha}\}$ defined by equation~\eqref{Eqn-CoefficientsNL}. An iteration reduces to two nested-loops: (i) compute from equation~\eqref{Eqn-ForceNL} all contact forces $P^{e_\alpha}$ associated with edges $e_\alpha$, (ii) update nonlocal terms $\gamma_{NL}^{e_\beta}$ with the contributions $w_{e_\alpha}^{e_\beta}$ induced in all incident contact pair-interactions $e_\beta$ (the right side of Figure~\ref{Fig-Algorithm-Incidence} illustrates all incident edges to edge $ab$). Finally, contact forces are determined from converged nonlocal contributions $\{\gamma_{NL}^{e_\alpha}\}$ and assembled into global forces $F$ (i.e., $F$ is a vector that collects the out-of-balanced force of each particle in the system). Thus, the nonlocal contact formulation only adds a computational overhead cost of complexity $O(\overline{n}N_c)$, where $N_c$ is the number of contact pair-interactions and $\overline{n}$ is the average number of neighbouring particles. The implementation is summarized in Algorithm~\ref{Alg-NonlocalAssembly}. It is worth noting that this is the only modification required for implementing the nonlocal contact formulation in existing codes.

\begin{algorithm}[htbp]
\caption{Nonlocal contact formulation. Computation of global forces.}
\label{Alg-NonlocalAssembly}
\begin{algorithmic}[1]
\REQUIRE $\{\gamma^{e_\alpha}\}$, TOL, and the simplicial complex $X$.
\STATE \textit{/$\ast$~~Compute nonlocal contributions $\{\gamma_{NL}^{e_\alpha}\}$~~$\ast$/}
\STATE $\text{Error} \leftarrow \text{TOL}$
\STATE $k \leftarrow 0$
\STATE $\{^0\gamma_{NL}^{e_\alpha}\} \leftarrow \{ 0 \}$
\WHILE{$\text{Error} \geq \text{TOL}$}
\STATE \textit{/$\ast$~~Loop over all contact pair-interactions $e_\alpha$~~$\ast$/}
\FOR  {$e_\alpha \in E_1(X)$}
\IF   {$\gamma^{e_\alpha} + {^k\gamma_{NL}^{e_\alpha}} > 0$}
\STATE \textit{/$\ast$~~Compute contact force from~\eqref{Eqn-ForceNL}~~$\ast$/}
\STATE $P^{e_\alpha} \leftarrow n_H^{e_\alpha} (\gamma^{e_\alpha} + {^k\gamma_{NL}^{e_\alpha}})^{3/2}$
\STATE \textit{/$\ast$~~Loop over all contact pair-interactions $e_\beta \in E_1(X)$ incident to $e_\alpha$~~$\ast$/}
\FOR  {$e_\beta \in$ {\sc Incident}$\left( e_\alpha \right)$}
\STATE \textit{/$\ast$~~Update $\gamma_{NL}^{e_\beta}$ with the nonlocal displacement \\
               \hspace{.21in} induced by contact force $P^{e_\alpha}$, i.e., $w_{e_\alpha}^{e_\beta}$ from \eqref{Eqn-NonlocalContribution}~~$\ast$/}
\STATE ${^{k+1}\gamma_{NL}^{e_\beta}} \leftarrow {^{k+1}\gamma_{NL}^{e_\beta}} + w_{e_\alpha}^{e_\beta}$
\ENDFOR
\ENDIF
\ENDFOR
\STATE \textit{/$\ast$~~Compute a measure of convergence~~$\ast$/}
\STATE $\text{Error} \leftarrow \| {^{k+1}\gamma_{NL}} - {^{k}\gamma_{NL}} \|$
\STATE $k \leftarrow k+1$
\ENDWHILE
\STATE \textit{/$\ast$~~Assemble global force $F$ using converged nonlocal terms $\{^k\gamma_{NL}^{e_\alpha}\}$~~$\ast$/}
\FOR  {$e_\alpha \in E_1(X)$}
\IF   {$\gamma^{e_\alpha} + {^k\gamma_{NL}^{e_\alpha}} > 0$}
\STATE $P^{e_\alpha} \leftarrow n_H^{e_\alpha} (\gamma^{e_\alpha} + {^k\gamma_{NL}^{e_\alpha}})^{3/2}$
\STATE $F \leftarrow$ {\sc Assemble}$\left( e_\alpha , P^{e_\alpha} \right)$
\ENDIF
\ENDFOR
\RETURN $F$
\end{algorithmic}
\end{algorithm}

\section{Validation of the nonlocal contact formulation}
\label{Section-Validation}

The validation of the nonlocal contact formulation presented in this work is twofold. First, we compare the analytical predictions with experimental results from Tatara \cite{Tatara-1989} for a rubber sphere pressed between two rigid plates. Second, we perform detailed finite-element simulations for different loading configurations and compare contact forces with predictions of the nonlocal formulation.

The experimental observations reported by Tatara \cite{Tatara-1989} correspond to rubber spheres of radius $R=0.01$~m and elastic properties $E=1.85$~MPa, $\nu=0.46$. The spheres were pressed between two rigid plates in a tensile-compression test machine and, by exploring deformations up to $40\%$ of the undeformed diameter, the author measured applied loads with a load-cell. Thus, applied load versus deformation curves were recorded. Tatara also reported that neither hysteresis nor permanent deformations were observed during the experiments.

\begin{figure}[htbp]
    \centering
    \includegraphics[scale=0.15]{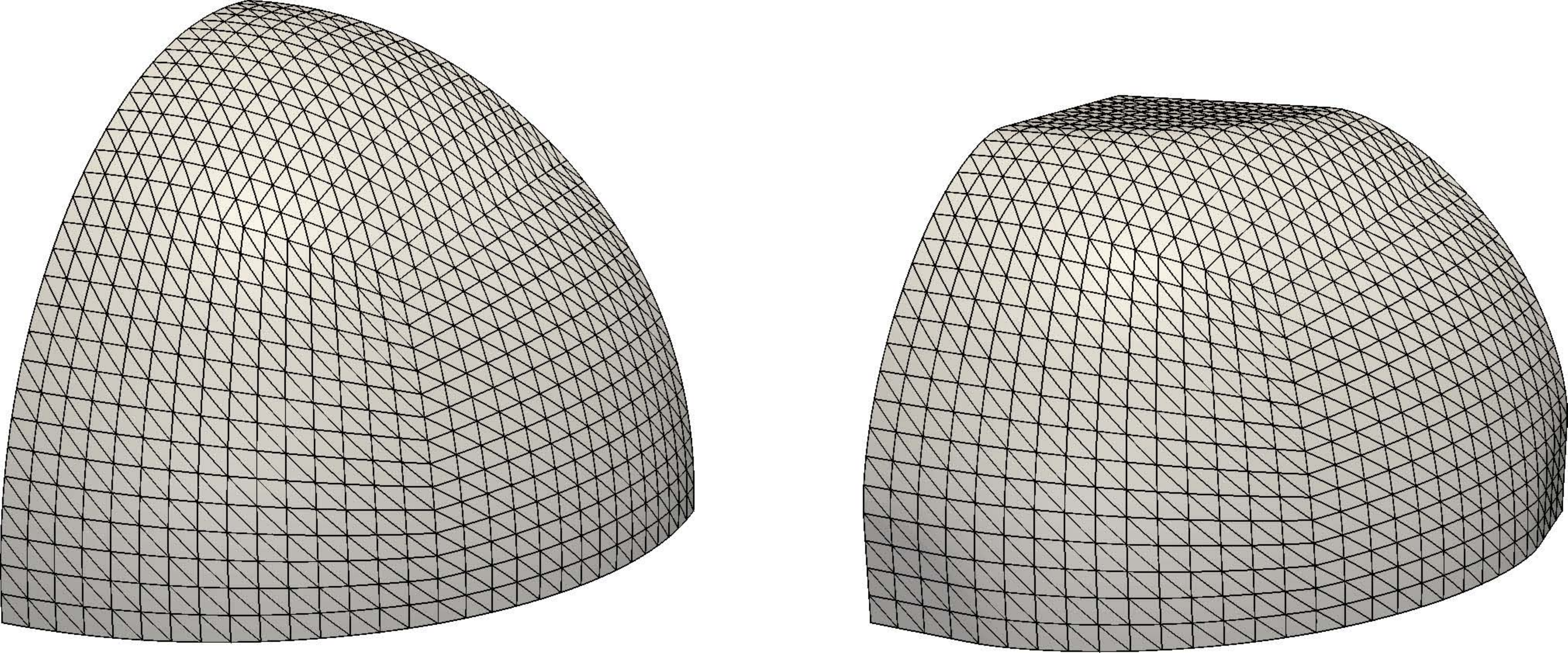}
    \caption{Finite-element mesh of one-eighth of a sphere---a coarse mesh comprised by 40,500 tetrahedral elements and 61,504 nodes is depicted. Left: Initial mesh. Right: Deformed mesh compressed in one direction and constrained in two other directions.}
    \label{Fig-TheMesh}
\end{figure}

A Lagrangian finite-element formulation is adopted and finite deformations are considered \cite{Zienkiewicz-2005}. Due to geometric and loading symmetries only one-eighth of the sphere is modeled. The mesh comprises 324,000 10-node tetrahedral isoparametric elements and 461,404 nodes. The three-dimensional solid is characterized by a strain-energy density of the form
\begin{equation}
    W(F) = \frac{1}{2} \left(\lambda - \frac{2}{3}\mu \right) \left( \log{J} \right)^2 - \mu \log{J} + \frac{\mu}{2} \text{tr}(\nabla{\varphi}^\text{T} \nabla{\varphi})
    \label{Eqn-NeohookeanStrainEnergy}
\end{equation}
which describes a neo-Hookean material extended to the compressible range. In this expression, $\varphi$ is the deformation mapping, $\nabla{\varphi}$ is the deformation gradient tensor, $J=\text{det}(\nabla{\varphi})$, and $\lambda$ and $\mu$ are the Lam\'{e} constants---the values employed in the simulations correspond to those reported by Tatara \cite{Tatara-1989}. Rigid frictionless boundary constraints are enforced by a penalization method, whereas symmetry conditions are imposed by restraining the corresponding degrees of freedom. The numerical solution is presumed ostensibly converged with respect to the mesh size---meshes with higher levels of refinement were used by way of reference. By way of illustration, Figure~\ref{Fig-TheMesh} shows a coarse mesh of one-eighth of a sphere.

\begin{figure}[htbp]
    \centering
    \begin{tabular}{cc}
    \includegraphics[scale=0.63]{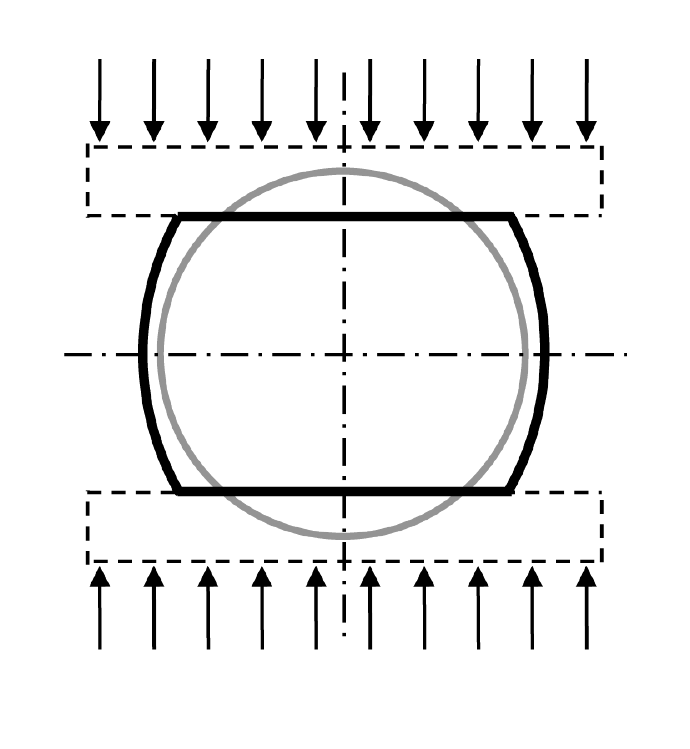}
    &
    \includegraphics[scale=0.56]{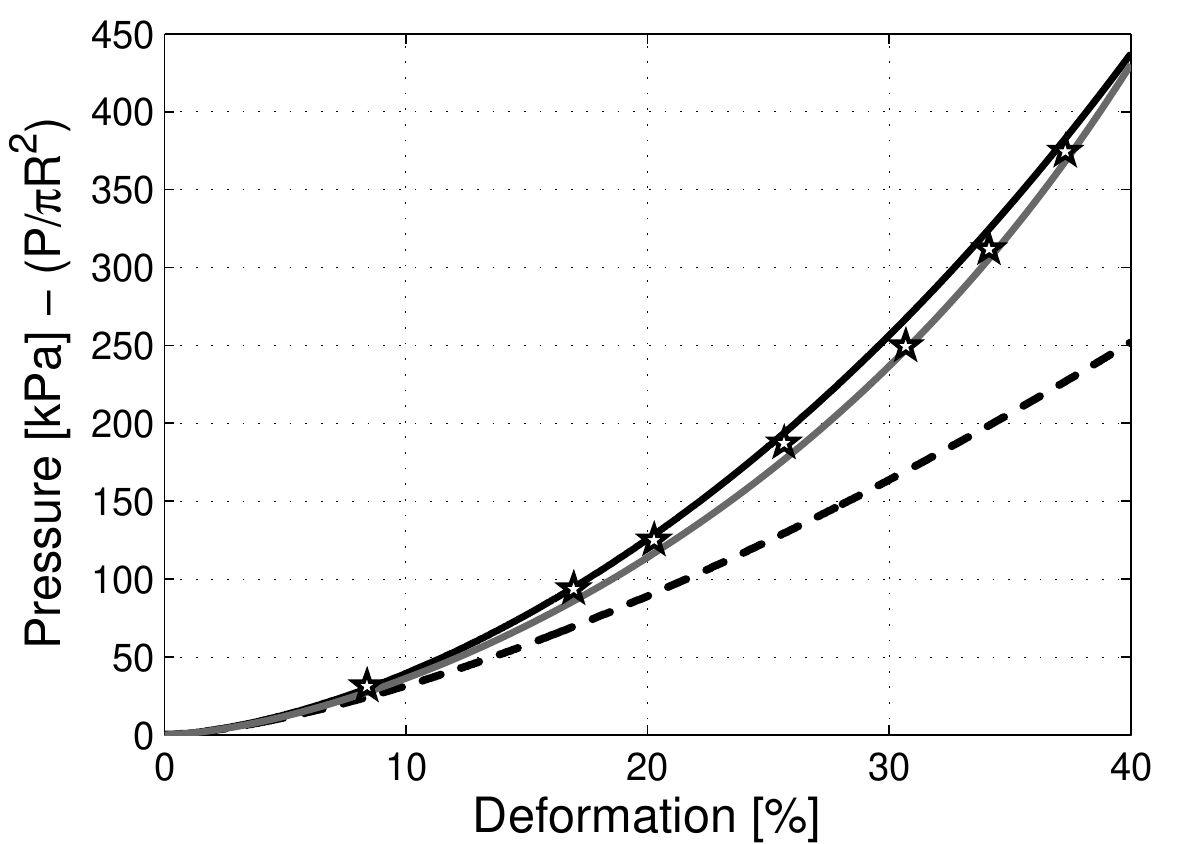}
    \end{tabular}
    \caption{Validation of the nonlocal contact formulation. Left: Schematic of the loading configuration. Right: Applied load versus deformation curves. Hertz theory prediction (black-dashed curve), nonlocal contact formulation results (black curve), finite element solution (grey curve), and experimental measurements from Tatara (1989) (five-pointed stars).}
    \label{Fig-StaticGC-Validation1}
\end{figure}

Figure~\ref{Fig-StaticGC-Validation1} summarizes the results of the validation in an applied-load-versus-deformation plot. Hertz theory predictions depart from experimental observations at very small deformations. In sharp contrast, the detailed finite-element simulation and the analytical predictions of the nonlocal contact formulation presented in this work are in excellent agreement with experimental measurements over the whole range of deformations. It bears emphasis that the computational effort required to evaluate a solution with the nonlocal formulation, or with Hertz theory, is negligible in comparison with the effort of solving the finite-element model.

\begin{figure}[htbp]
    \centering
    \begin{tabular}{cc}
    \includegraphics[scale=0.63]{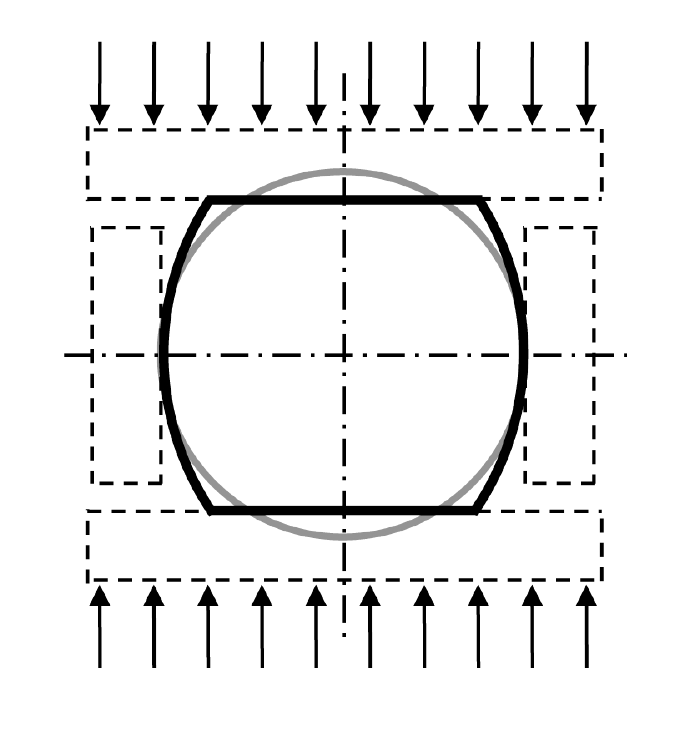}
    &
    \includegraphics[scale=0.56]{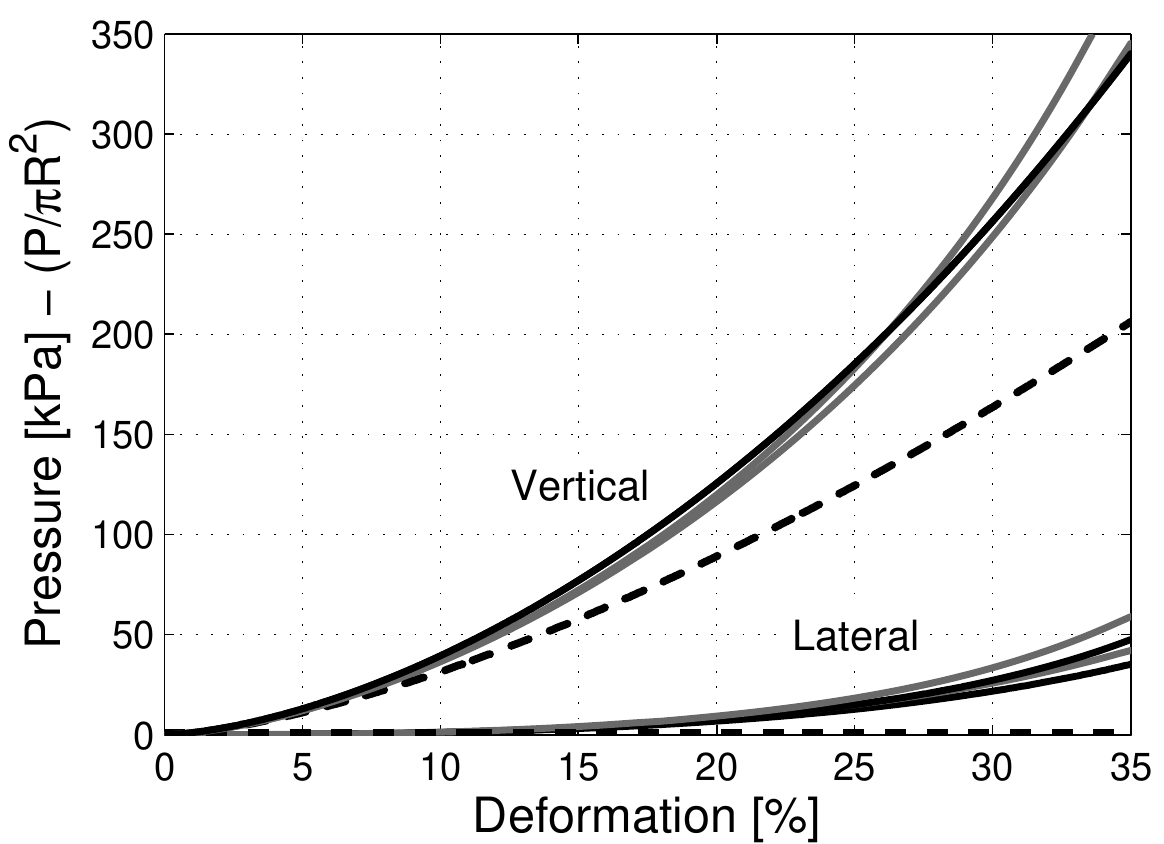}
    \end{tabular}
    \caption{Validation of the nonlocal contact formulation. Left: Schematic of a spherical particle compressed in one direction and constrained in other direction. Right: Applied load and lateral reaction versus deformation curves for a spherical particle compressed in one direction and constrained in one/two directions. Hertz theory prediction (black-dashed curve), nonlocal contact formulation results (black curve), and finite element solution (grey curve). Both loading configurations are intentionally depicted with the same color scheme.}
    \label{Fig-StaticGC-Validation2}
\end{figure}

In order to asses the accuracy of the nonlocal contact formulation in configurations where nonlocal mesoscopic deformations play an even larger role, we study a spherical elastic particle compressed in one direction and constrained in other directions. Due to the lack of experimental observations, we perform detailed finite-element simulations and compare contact forces with predictions of the nonlocal formulation. Specifically, Figure~\ref{Fig-StaticGC-Validation2} shows the applied load and the lateral reactions versus deformation of configurations with one set of lateral constraints and with two orthogonal sets of lateral constraints. The lateral reaction is a consequence of the lateral geometric confinement of the systems. Hertz theory entirely disregards such effect and predicts no lateral reaction forces. The present formulation, in contrast, is designed to account for nonlocal mesoscopic-deformation effects and therefore lateral reactions naturally result from the compressive loading conditions. The agreement between predictions of the finite-element model and the nonlocal contact formulation is again remarkable.

It bears emphasis that despite the good agreement at large effective deformations, the extension of the nonlocal contact formulation to nonlinear material models and to finite deformations remains a topic of interest.

\section{Macroscopic behavior of confined granular systems}
\label{Section-MacroBehavior}

\subsection{Granular crystals}

\begin{figure}[htbp]
    \centering
    \includegraphics[scale=0.52]{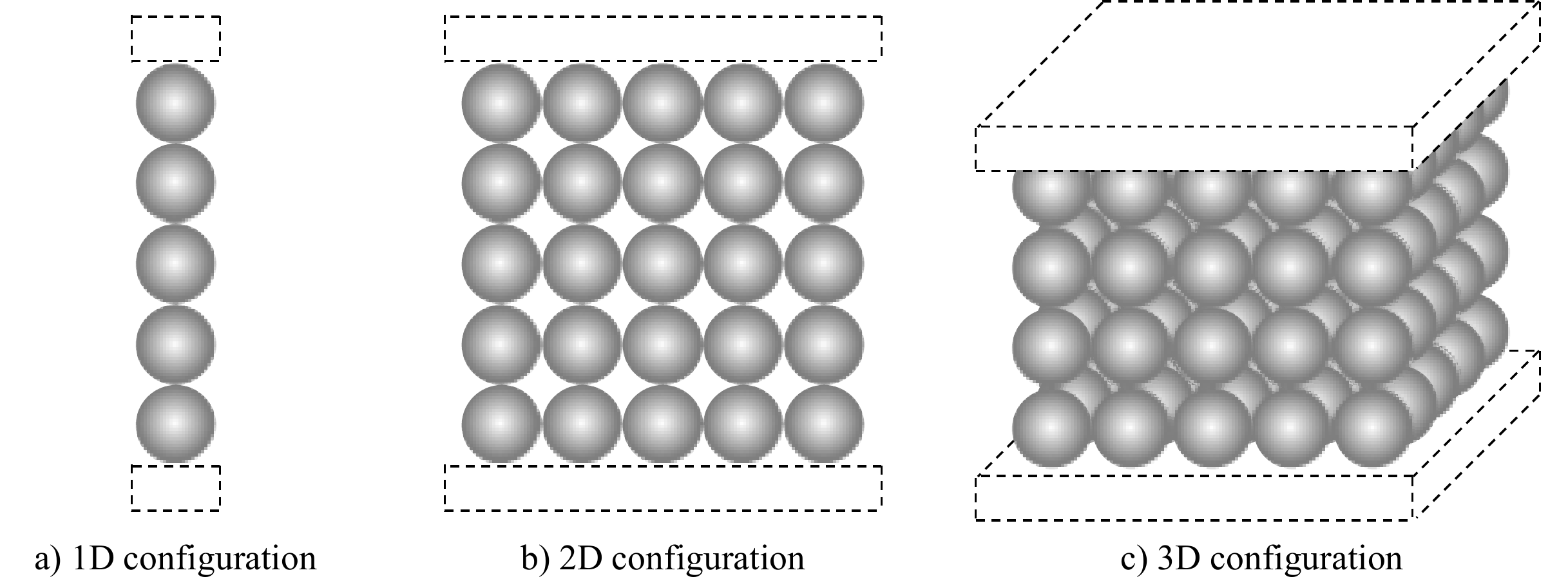}
    \caption{Different granular crystals assemblies. The bi-dimensional and three-dimensional configurations are periodic in all directions orthogonal to the load.}
    \label{Fig-ConfigGC}
\end{figure}

Granular crystals, or highly packed granular lattices, allow for a systematic understanding of nonlocal mesoscopic-deformation effects in the macroscopic behavior of granular systems. Specifically, we study different dimensional configurations, different packings and different volumetric confinements of periodic homogeneous systems, i.e., all particles are of same size and material properties. By way of example, and for later reference, we illustrate in Figure~\ref{Fig-ConfigGC} a one-dimensional chain, a bi-dimensional square lattice, and a three-dimensional cubic lattice of spherical beads. It is worth noting that the loading configurations discussed in Section~\ref{Section-Validation} are equivalent to the granular crystals depicted in Figure~\ref{Fig-ConfigGC}. The experimental setup is equivalent to the one-dimensional configuration, and the configurations constrained in one and two orthogonal directions correspond to bi-dimensional and three-dimensional arrangements, respectively.

\begin{figure}[htbp]
    \centering
    \begin{tabular}{cc}
    \includegraphics[scale=0.56]{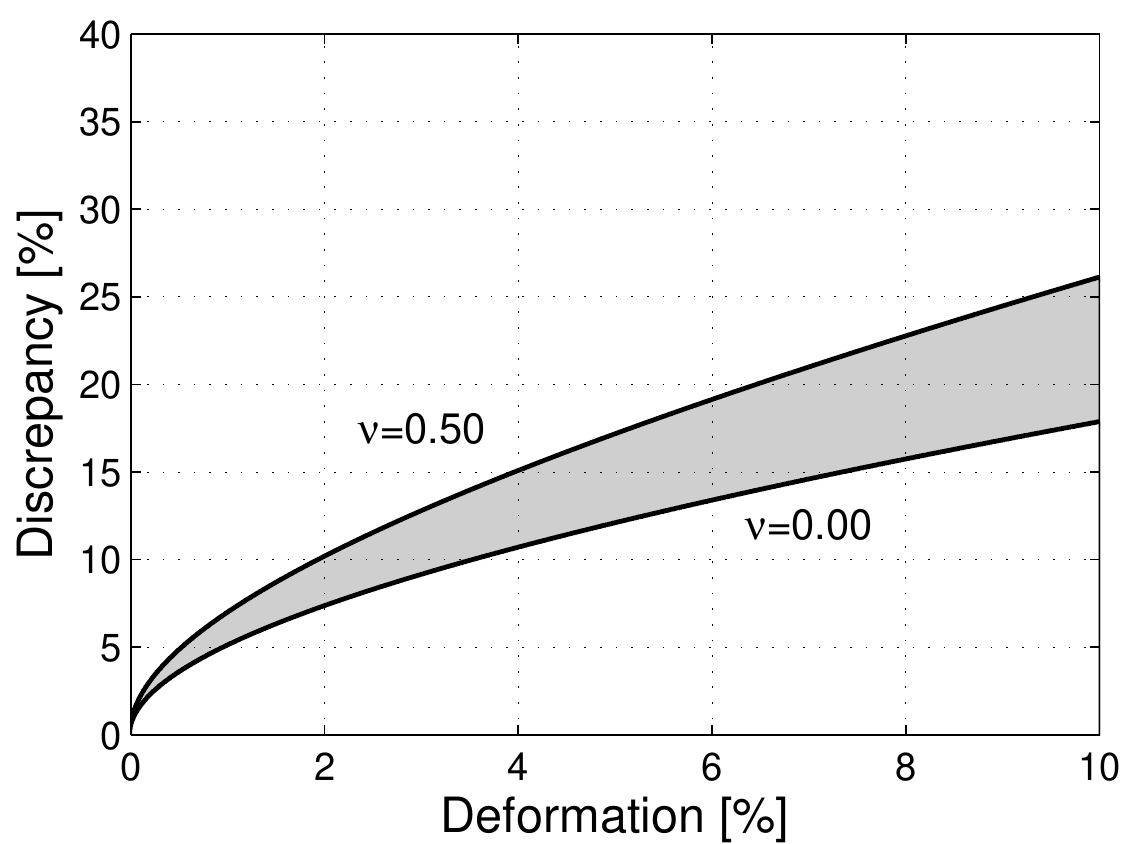}
    &
    \includegraphics[scale=0.56]{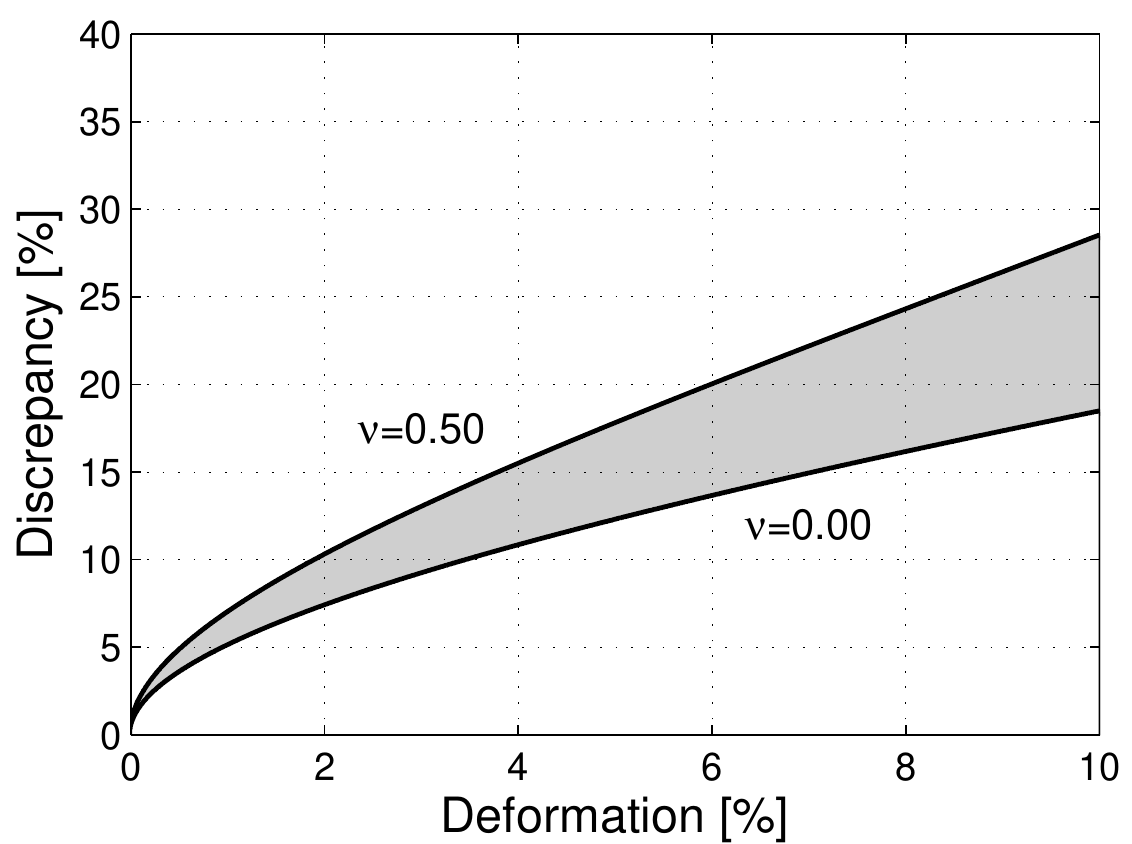}
    \end{tabular}
    \caption{Discrepancy between the nonlocal contact formulation and Hertz theory for $\nu\in(0,0.50)$. Left: One-dimensional configuration (Figure~\ref{Fig-ConfigGC}a). Right: Three-dimensional configuration (Figure~\ref{Fig-ConfigGC}c).}
    \label{Fig-ErrorGC-nD}
\end{figure}

The discrepancy between Hertz theory and the extended theory presented in this work evidently depends on the granular assembly and on the loading conditions. Hence, we restrict attention to deformations smaller than $10\%$ and we quantify the discrepancy between the two formulations by
\begin{equation}
    \mathcal{D}
    =
    \frac{P(\gamma)-n_H\gamma^{3/2}}{n_H\gamma^{3/2}}
\end{equation}
where $n_H\gamma^{3/2}$ is the prediction of Hertz theory and $P(\gamma)$ corresponds to the result of the nonlocal theory, i.e., the solution of Equations~\eqref{Eqn-ForceNL} and \eqref{Eqn-CoefficientsNL}. Then, for a one-dimensional configuration, Hertz theory differs from the nonlocal contact formulation by
\begin{equation}
    \mathcal{D}_{\text{1D}}
    =
    \frac{1}{2\pi}\left(\frac{3-2\nu}{1-\nu}\right)\epsilon^{1/2} +
    \frac{7}{24\pi^2}\left(\frac{3-2\nu}{1-\nu}\right)^2\epsilon +
    \mathcal{O}(\epsilon^{3/2})
\end{equation}
where $\epsilon=\frac{\gamma}{2R}$. For bi-dimensional and three-dimensional configurations, the biaxial and triaxial loading conditions introduce a new term of order $\mathcal{O}(\epsilon^{5/4})$ and therefore the first two terms in the error expansion are unperturbed, i.e.,
\begin{equation}
    \mathcal{D}_{\text{2D,3D}}
    =
    \frac{1}{2\pi}\left(\frac{3-2\nu}{1-\nu}\right)\epsilon^{1/2} +
    \frac{7}{24\pi^2}\left(\frac{3-2\nu}{1-\nu}\right)^2\epsilon +
    \mathcal{O}(\epsilon^{5/4})
\end{equation}
It bears emphasis that the discrepancy between the nonlocal contact formulation and Hertz theory depends strongly on the deformation $\epsilon$ and the Poisson's ratio $\nu$, regardless the value of the Young modulus and the size of the particles. It also depends weakly on the dimensional configuration of the granular crystal. This last statement is in agreement with the numerical results presented in Figure~\ref{Fig-StaticGC-Validation2}. The error bounds presented above are illustrated in Figure~\ref{Fig-ErrorGC-nD} for $\nu\in(0,0.50)$.

\begin{figure}[htbp]
    \centering
    \begin{tabular}{cc}
    \includegraphics[scale=0.49]{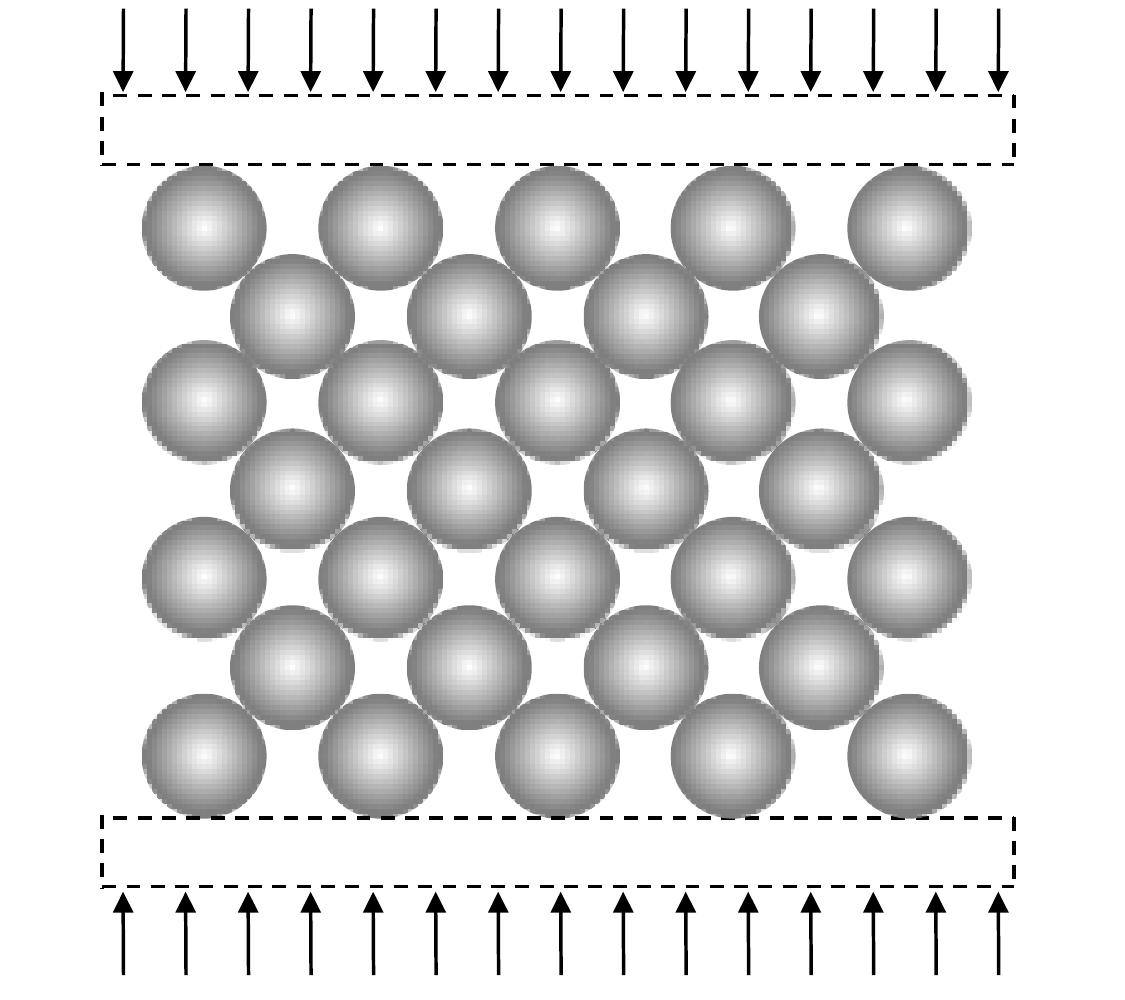}
    &
    \includegraphics[scale=0.56]{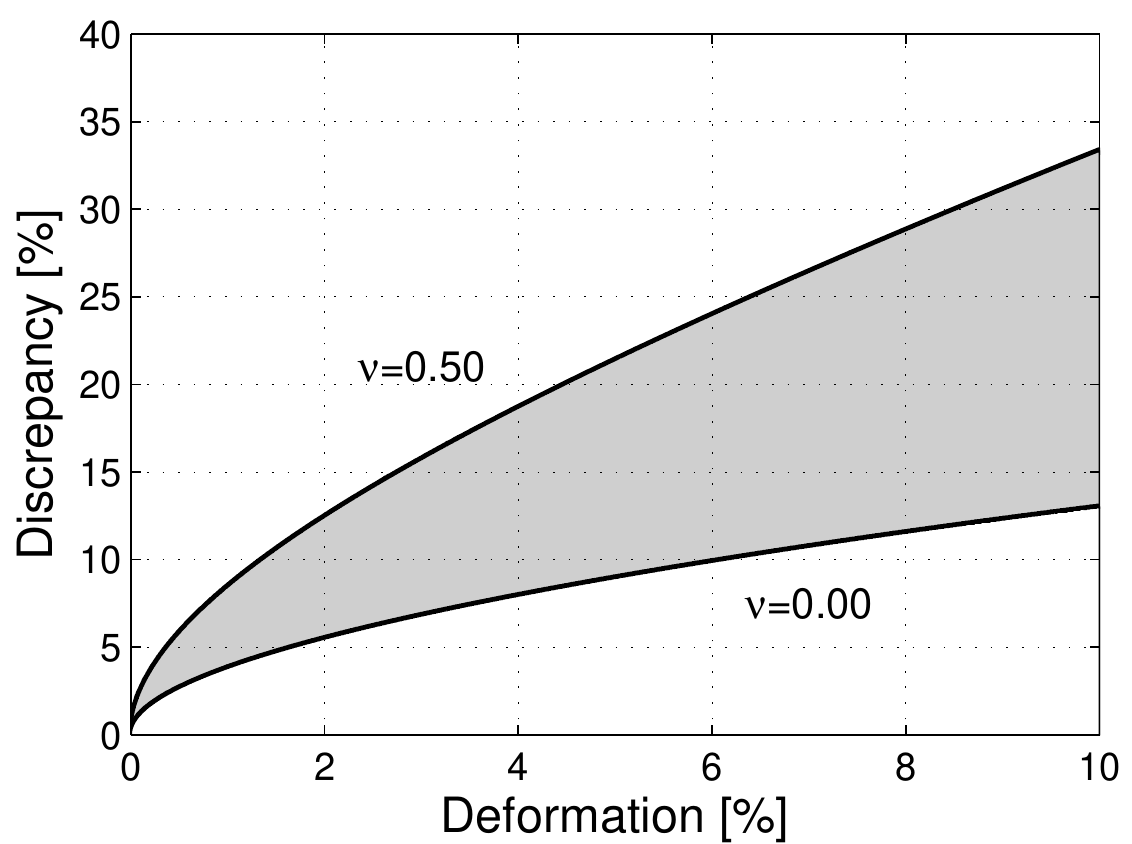}
    \end{tabular}
    \caption{Left: Bi-dimensional granular crystal rotated by $\pi/4$---the configuration is periodic in the direction orthogonal to the load. Right: Discrepancy between the nonlocal contact formulation and Hertz theory for $\nu\in(0,0.50)$.}
    \label{Fig-ConfigGC-Rotated}
\end{figure}

\begin{figure}[htbp]
    \centering
    \begin{tabular}{cc}
    \includegraphics[scale=0.49]{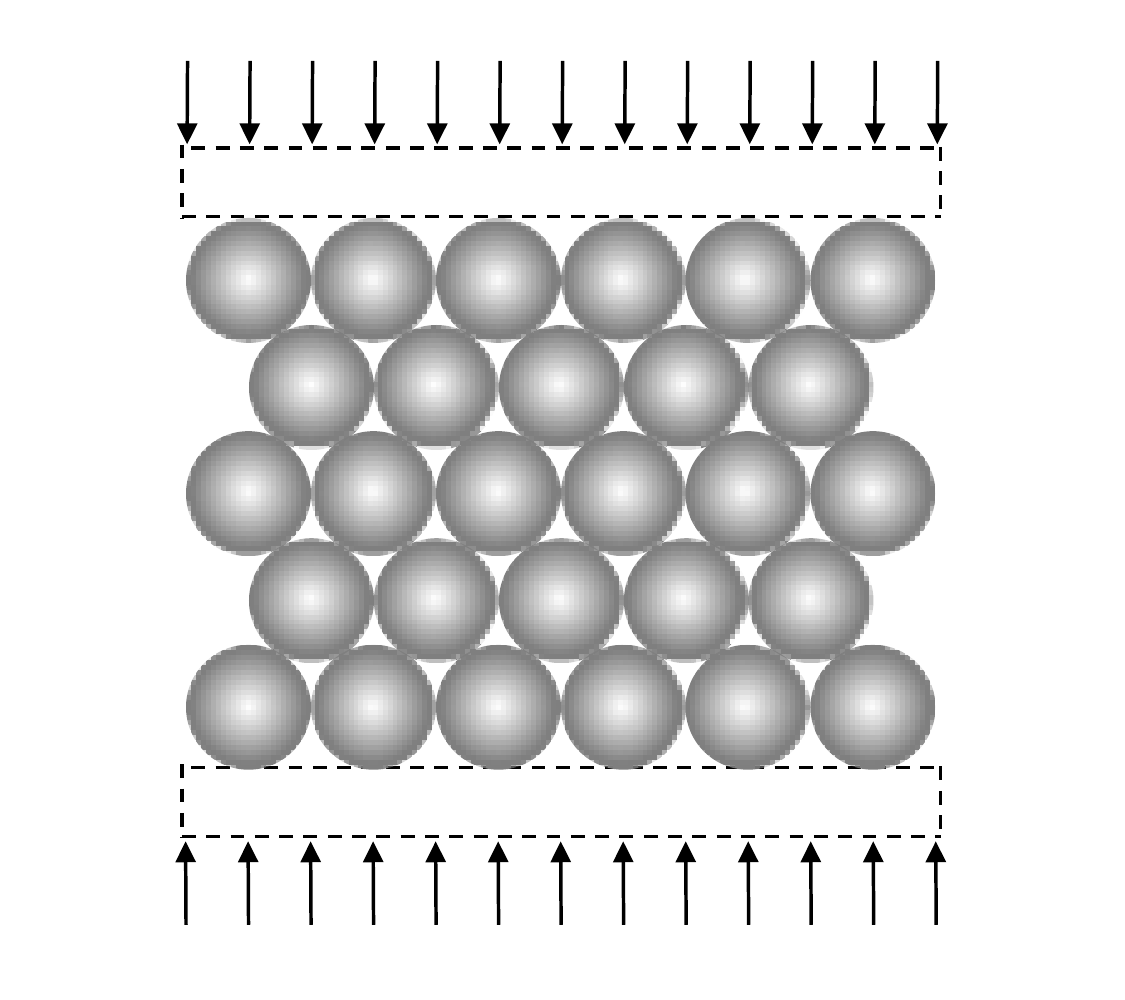}
    &
    \includegraphics[scale=0.56]{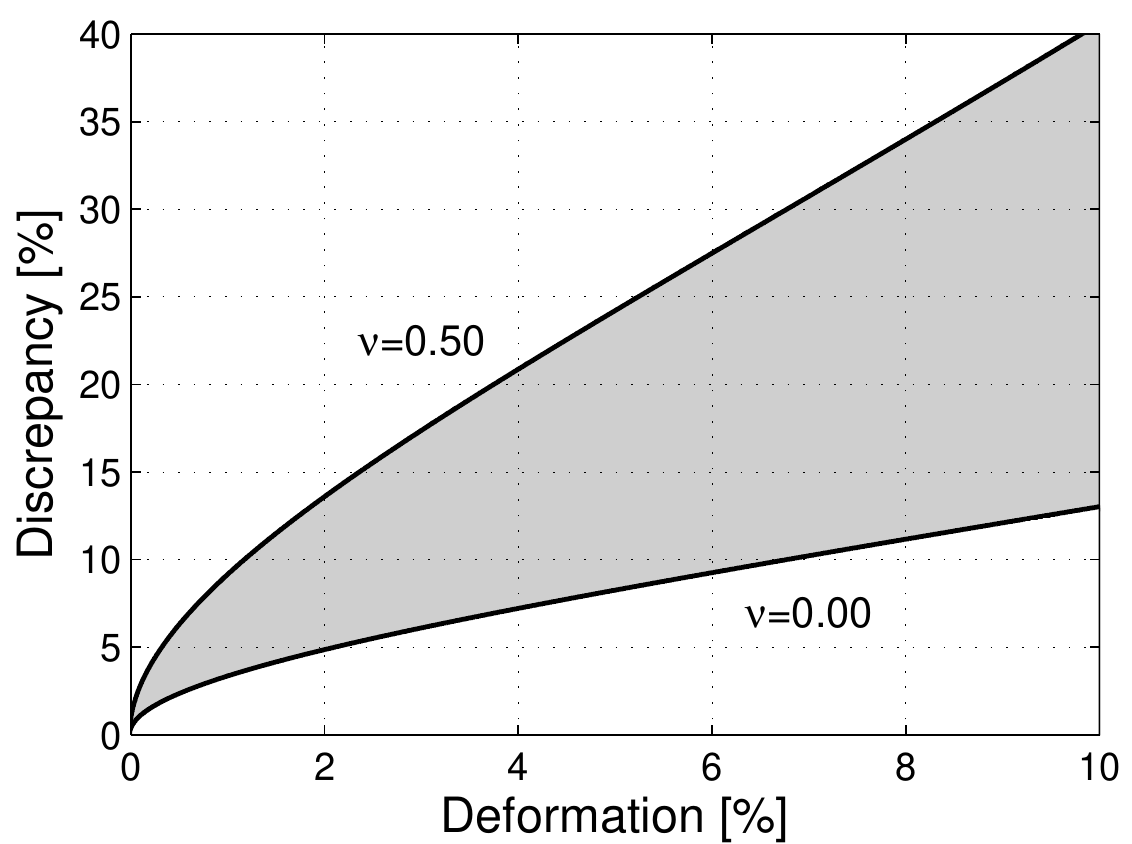}
    \end{tabular}
    \caption{Left: Close-packed bi-dimensional granular crystal---the configuration is periodic in the direction orthogonal to the load. Right: Discrepancy between the nonlocal contact formulation and Hertz theory for $\nu\in(0,0.50)$.}
    \label{Fig-ConfigGC-Packed}
\end{figure}

\begin{figure}[htbp]
    \centering
    \begin{tabular}{cc}
    \includegraphics[scale=0.49]{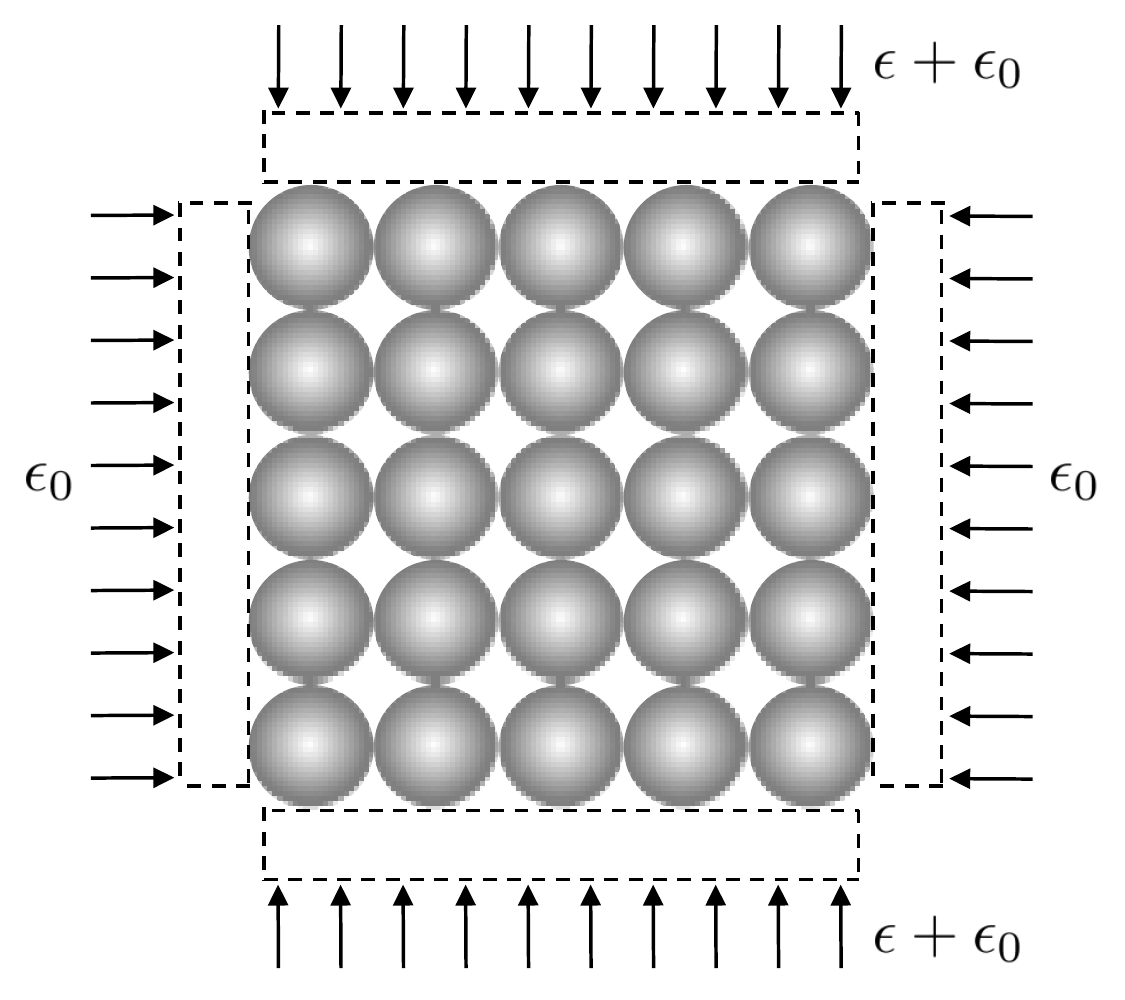}
    &
    \includegraphics[scale=0.56]{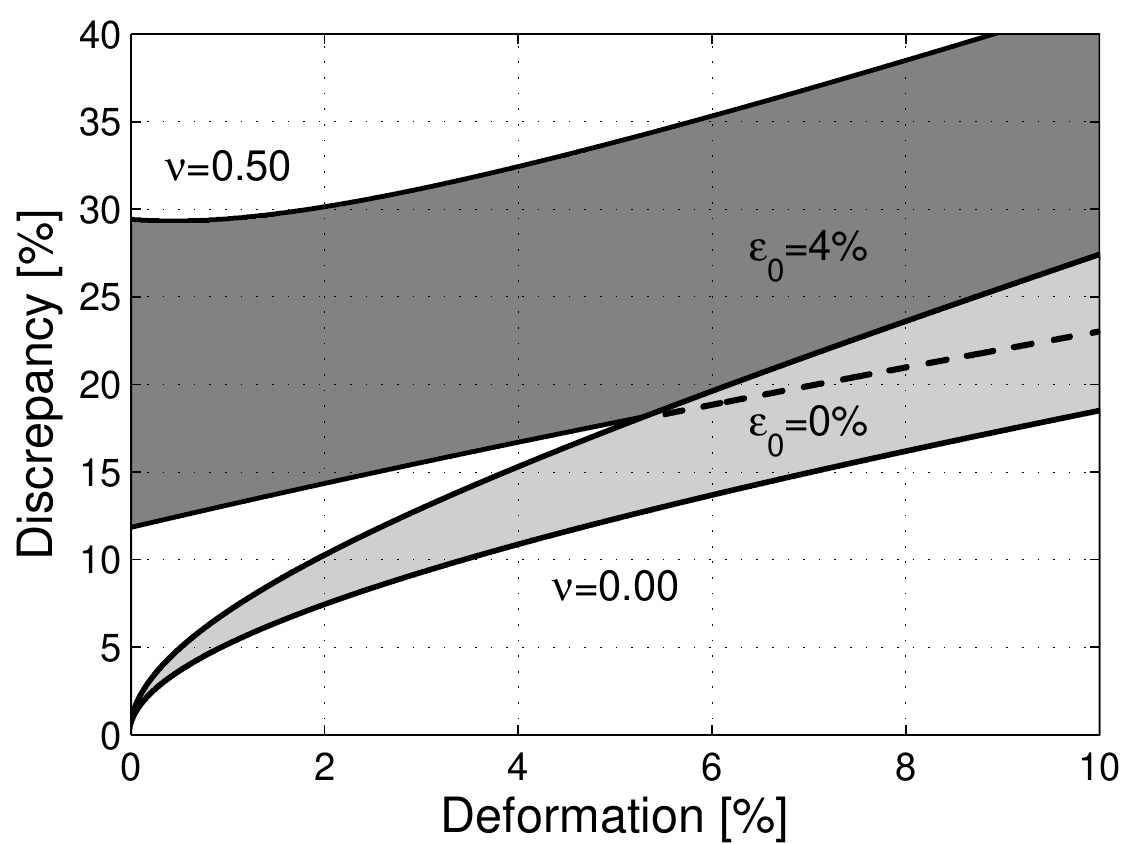}
    \end{tabular}
    \caption{Left: Bi-dimensional granular crystal preloaded by $\epsilon_0$ in both directions. Right: Discrepancy between the nonlocal contact formulation and Hertz theory for $\epsilon_0 = 4\%$ and $\nu\in(0,0.50)$.}
    \label{Fig-ConfigGC-Preloaded}
\end{figure}

The weak dependency on the dimensional configuration of the granular crystal suggests the study of square and cubic lattices loaded in different directions, e.g., the configuration depicted in Figure~\ref{Fig-ConfigGC-Rotated}, and of different assemblies, e.g., the close-packed configuration depicted in Figure~\ref{Fig-ConfigGC-Packed}. Similarly, it is interesting to investigate the role of initial confinement on the macroscopic behavior of the system, e.g., the preloaded configuration depicted in Figure~\ref{Fig-ConfigGC-Preloaded}. These three cases are examined next in turn. For a displacement $\gamma$ in the direction of loading, deformations $\epsilon$ of rotated and close-packed configurations are defined by $\epsilon = \frac{\gamma}{\sqrt{2}R}$ and $\epsilon = \frac{\gamma}{\sqrt{3}R}$, respectively.

We note from Figures~\ref{Fig-ErrorGC-nD} and \ref{Fig-ConfigGC-Rotated} that nonlocal mesoscopic-deformations in lattices of incompressible materials (i.e., $\nu \rightarrow 1/2$) are more relevant for rotated configurations. However, highly compressible materials (i.e., $\nu \rightarrow 0$) are less sensible to nonlocal effects in rotated square lattice assemblies. Granular crystals comprised by rotated square or cubic lattices are difficult to realize experimentally and therefore close-packed configurations become a more attractive approach. Figure~\ref{Fig-ConfigGC-Packed} shows that the discrepancy between the nonlocal contact formulation and Hertz theory is again larger for incompressible materials and smaller for highly compressible particles, when compared with the previous configurations. It is also interesting to observe that the nonlocal formulation predicts the formation of a gap between horizontal neighbors of close-packed configurations of highly compressible beads.

\begin{figure}[htbp]
    \centering
    \includegraphics[scale=0.56]{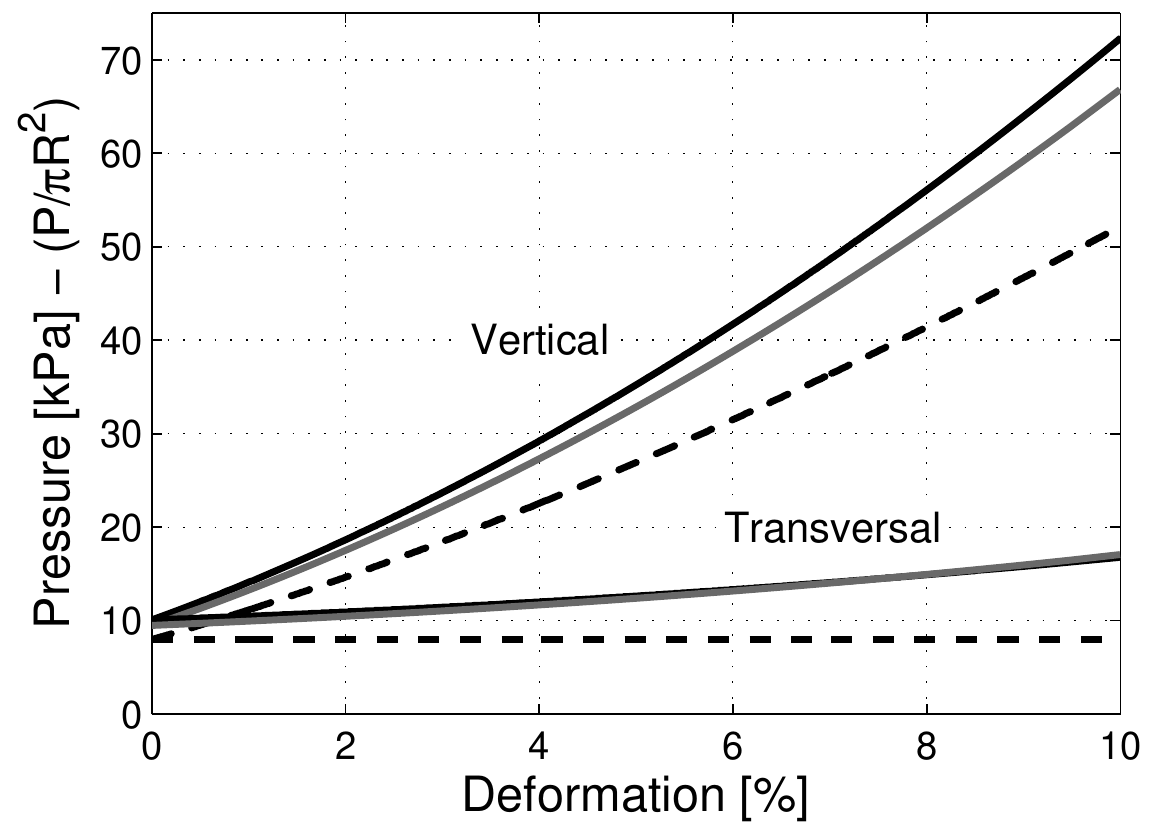}
    \caption{Applied load, and lateral reaction, versus deformation for a bi-dimensional granular crystal preloaded by $\epsilon_0 = 4\%$ configuration. Hertz theory prediction (black-dashed curve), nonlocal contact formulation results (black curve), and finite element solution (grey curve).}
    \label{Fig-StaticGC-Validation3}
\end{figure}

The effect of volumetric confinement is shown in Figure~\ref{Fig-ConfigGC-Preloaded}. In order to confine a bi-dimensional granular crystal by $\epsilon_0$ in both directions, an initial load has to be applied to the system. Therefore, if the confined configuration is taken as the reference configuration, discrepancies between the nonlocal contact formulation and Hertz theory will be significant even for $\epsilon = 0$. Moreover, and in contrast to all previous configurations, it is evident in the figure that the error is not order $\mathcal{O}(\epsilon^{1/2})$ at small deformations. This different behavior suggests an additional validation of the nonlocal formulation with a detailed finite-element simulation. Figure~\ref{Fig-StaticGC-Validation3} shows the applied load, and lateral reaction, versus deformation curves for a bi-dimensional system with an initial confinement of $\epsilon_0 = 4\%$ and subsequent compression in one direction. The agreement between the detailed finite-element simulation and the analytical predictions of the nonlocal contact formulation is remarkable.

\begin{figure}[htbp]
    \centering
    \includegraphics[scale=0.49]{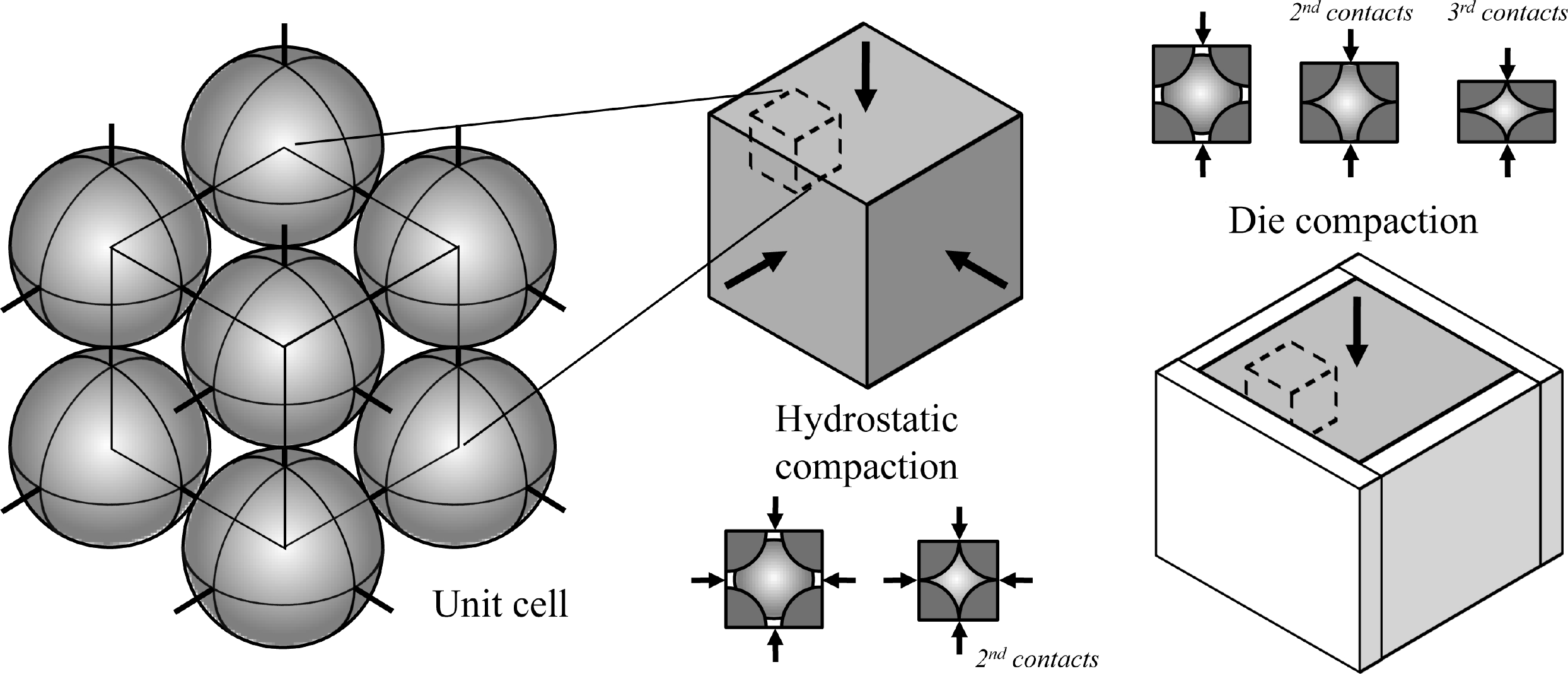}
    \caption{Body centered cubic granular crystal under hydrostatic compaction and die compaction.}
    \label{Fig-ConfigGC-BCC}
\end{figure}

Finally, we study the compaction of a three-dimensional configuration. Specifically, the behavior of a body centered cubic granular crystal under hydrostatic and die compaction is investigated (see Figure~\ref{Fig-ConfigGC-BCC}). The relative density is adopted as measure of deformation, that is
\begin{equation}
    \rho_R
    =
    \frac{\pi \sqrt{3}}{8(1-\epsilon_x)(1-\epsilon_y)(1-\epsilon_z)}
\end{equation}
where $\epsilon_x$, $\epsilon_y$, $\epsilon_z$ are the deformations of the unit cell in each coordinate direction (e.g., for hydrostatic compaction $\epsilon_x=\epsilon_y=\epsilon_z$, and for die compaction $\epsilon_x=\epsilon_y=0$ with $\epsilon_z\ne0$). Then, the undeformed system has $\rho_R=0.68$. It is worth noting that second contacts are activated during the compaction of a body centered cubic configuration---indeed, for die compaction, third contacts are also activated at large relative densities. This phenomenon results in an increasingly stiffer system and therefore large loads are required to achieve high-density compaction. Evidently, the accurate prediction of the onset of second and third contacts relies on the ability of the model to account for nonlocal mesoscopic-deformation effects.

\begin{figure}[htbp]
    \centering
    \begin{tabular}{cc}
    \includegraphics[scale=0.56]{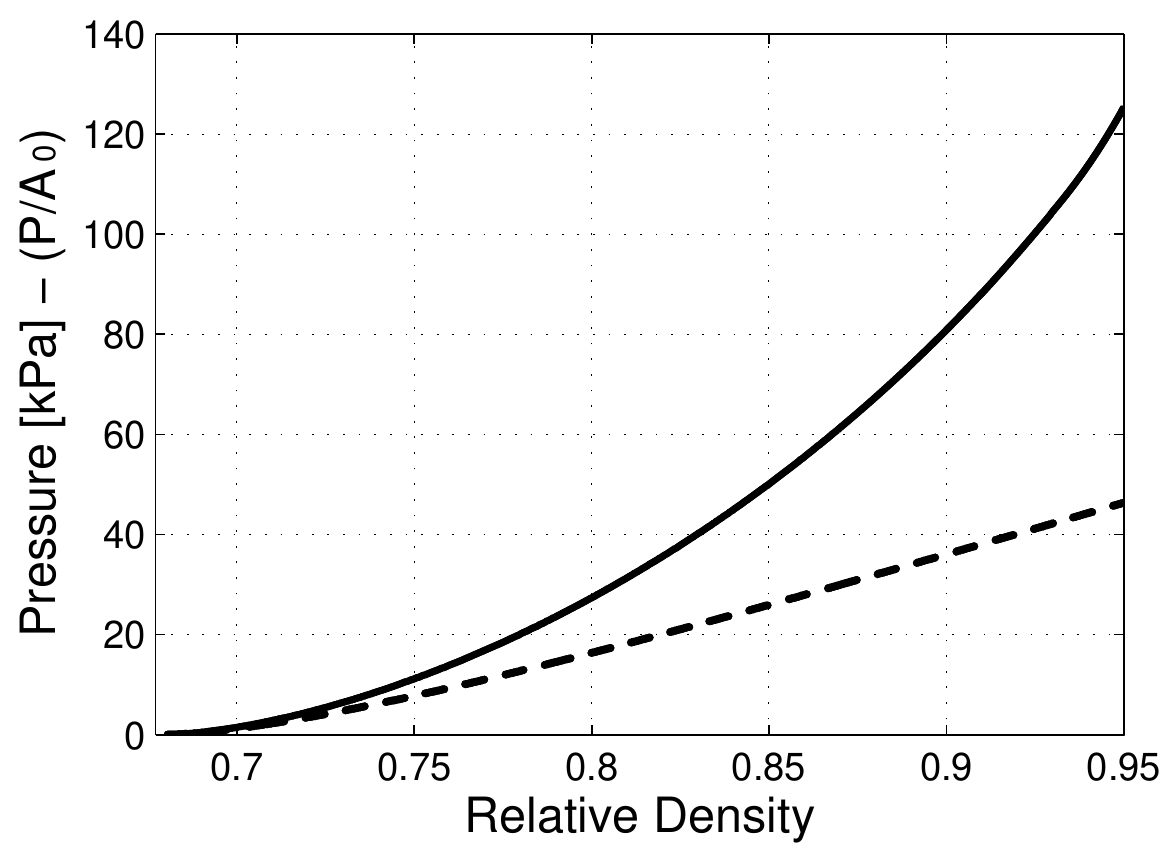}
    &
    \includegraphics[scale=0.56]{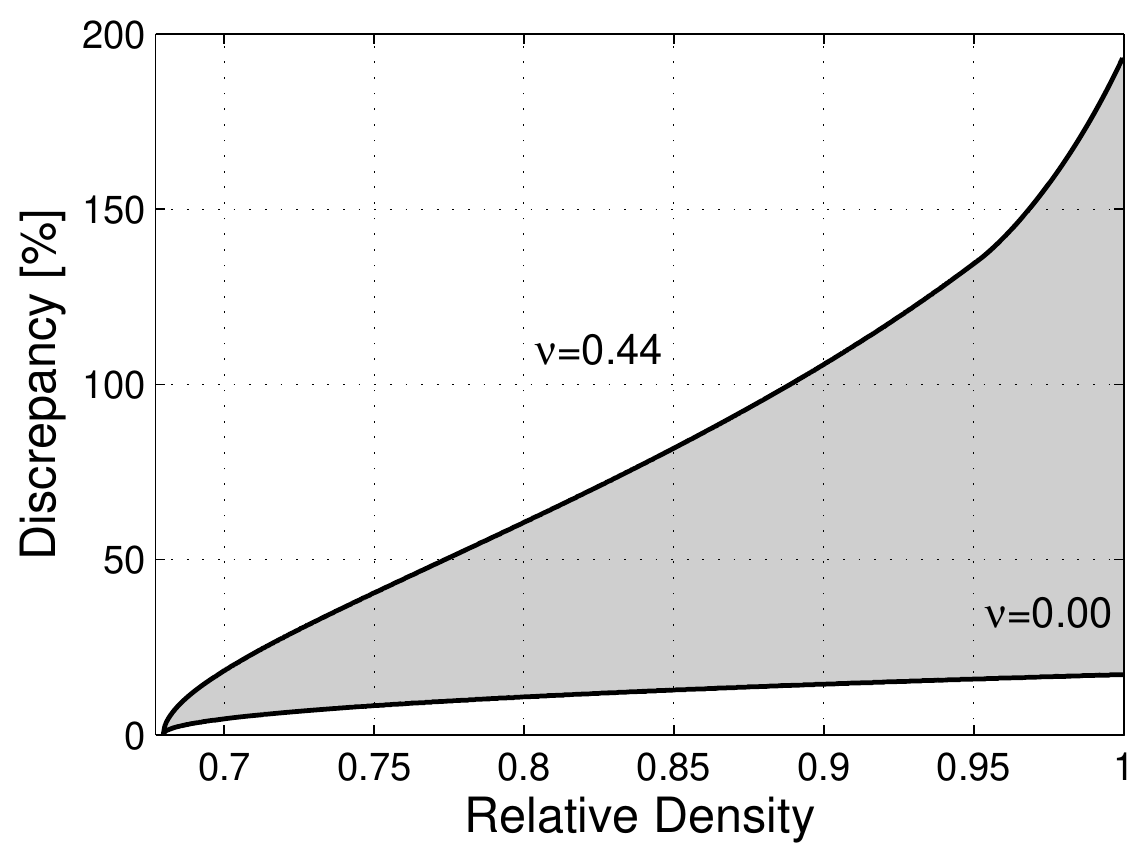}
    \end{tabular}
    \caption{Body centered cubic granular crystal under hydrostatic compaction. Left: Applied load versus relative density predicted by Hertz theory (black-dashed curve) and by the nonlocal contact formulation (black curve)---$R=0.01$~m, $E=1.85$~MPa, $\nu=0.46$. Right: Discrepancy between the nonlocal contact formulation and Hertz theory for $\nu\in(0,0.44)$.}
    \label{Fig-ConfigGC-BCC-ResultsHydro}
\end{figure}

The right side of Figure~\ref{Fig-ConfigGC-BCC-ResultsHydro} shows the discrepancy between the nonlocal contact formulation and Hertz theory in the prediction of the loading force required to achieve hydrostatic compaction of the granular system. For highly incompressible materials (i.e., $\nu \rightarrow 1/2$), it is interesting to observe that the discrepancy first changes its curvature and then rapidly grows for relative densities larger than $0.95$. This result is explained by the formation of second contacts. The nonlocal formulation predicts the formation of second contacts at $\rho_R=0.95$ for $\nu=0.45$, whereas Hertz theory at $\rho_R=1.04$ for all Poisson's ratios. The left side of the figure shows the applied load versus relative density obtained with both formulations for a body centered cubic packing of rubber spheres of radius $R=0.01$~m and elastic properties $E=1.85$~MPa, $\nu=0.46$. The discrepancy between the nonlocal formulation and Hertz theory is evident in the figure.

\begin{figure}[htbp]
    \centering
    \begin{tabular}{cc}
    \includegraphics[scale=0.56]{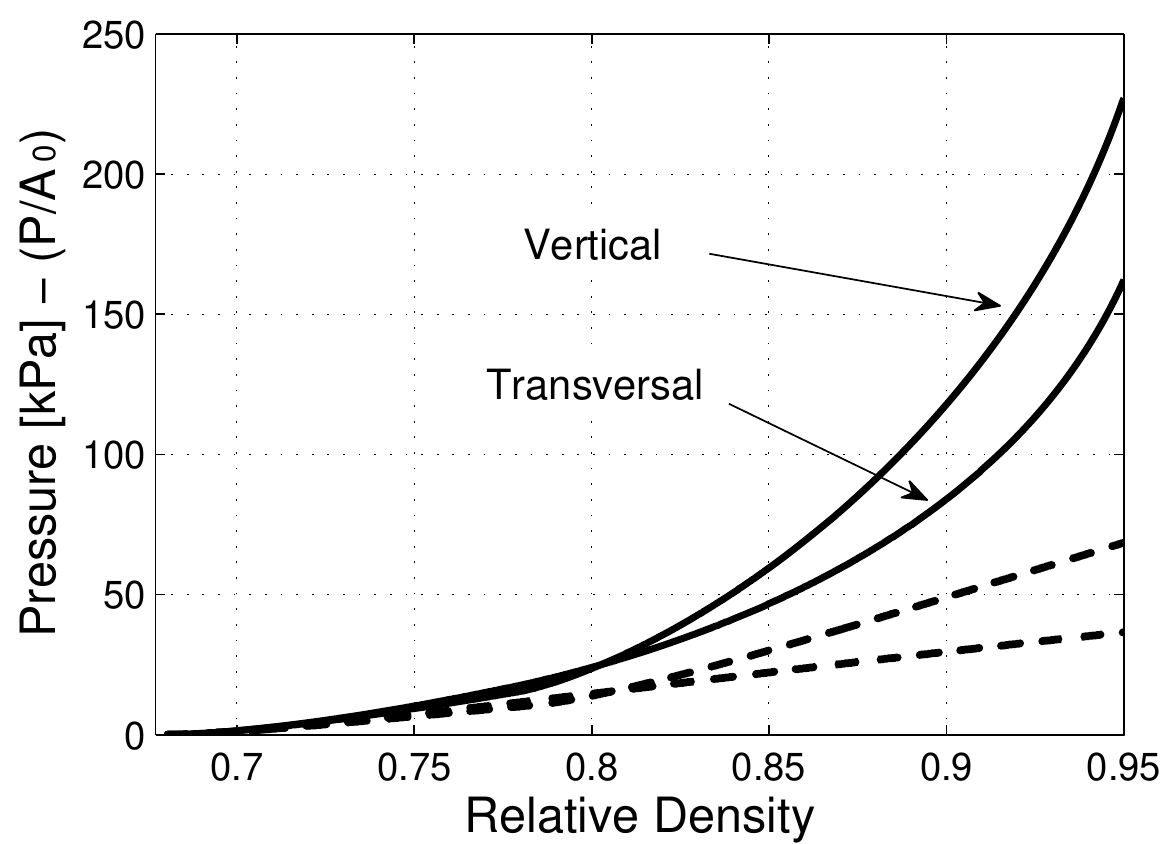}
    &
    \includegraphics[scale=0.56]{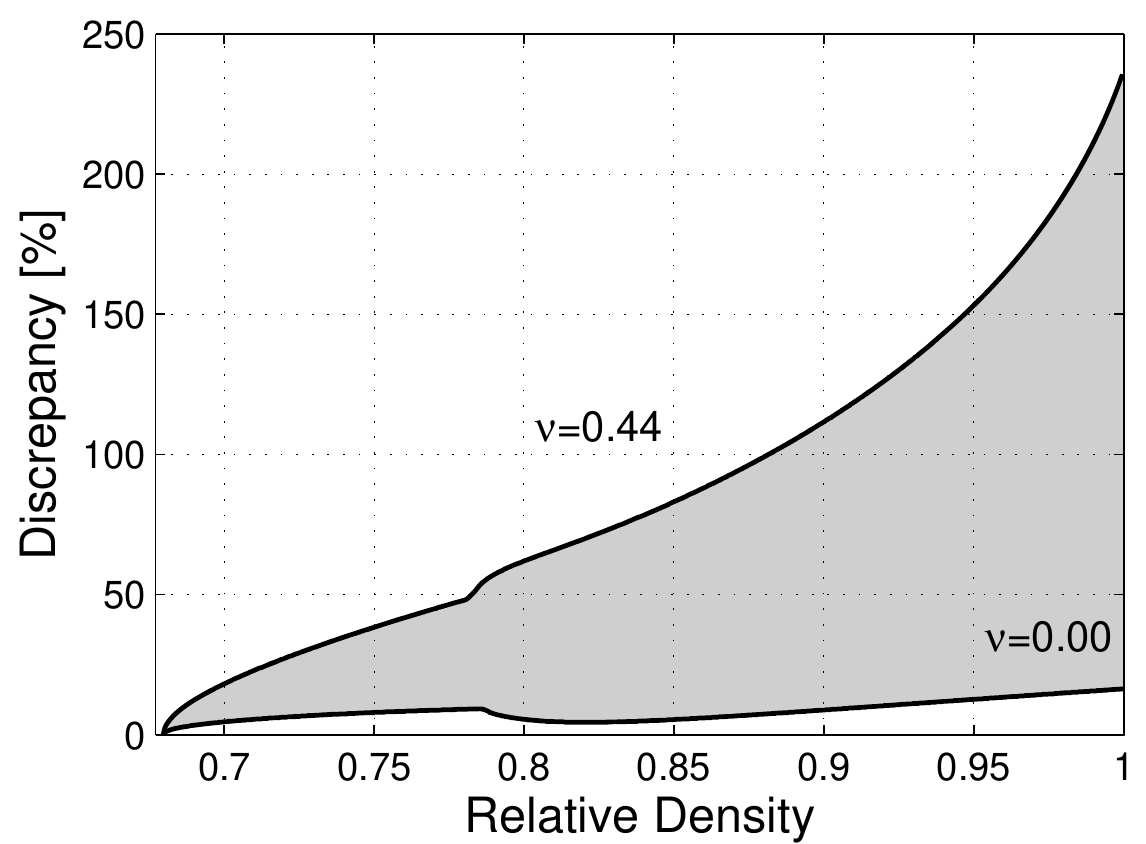}
    \end{tabular}
    \caption{Body centered cubic granular crystal under die compaction. Left: Applied load, and lateral reaction, versus relative density predicted by Hertz theory (black-dashed curve) and by the nonlocal contact formulation (black curve)---$R=0.01$~m, $E=1.85$~MPa, $\nu=0.46$. Right: Discrepancy between the nonlocal contact formulation and Hertz theory for $\nu\in(0,0.44)$.}
    \label{Fig-ConfigGC-BCC-ResultsDie}
\end{figure}

The right side of Figure~\ref{Fig-ConfigGC-BCC-ResultsDie} shows the discrepancy between the nonlocal contact formulation and Hertz theory in the prediction of the loading force required to achieve die compaction of the granular system. It is worth noting the presence of a corner point for values of relative density in the vicinity of $0.78$. Again, this behavior is explained by the formation of second contacts. The nonlocal formulation predicts the formation of second contacts at $\rho_R=0.780$ for $\nu=0.45$, whereas Hertz theory at $\rho_R=0.786$ for all Poisson's ratios. The left side of the figure shows the applied load, and lateral reaction, versus relative density obtained with both formulations for a body centered cubic packing of rubber spheres of radius $R=0.01$~m and elastic properties $E=1.85$~MPa, $\nu=0.46$. In this example, the nonlocal formulation predicts the formation of third contacts at $\rho_R=0.96$. As it was for the previous example, the discrepancy between both formulations is evident in the figure.

\subsection{Disordered granular media}

\begin{figure}[htbp]
    \centering
    \begin{tabular}{ccc}
    \includegraphics[scale=0.45]{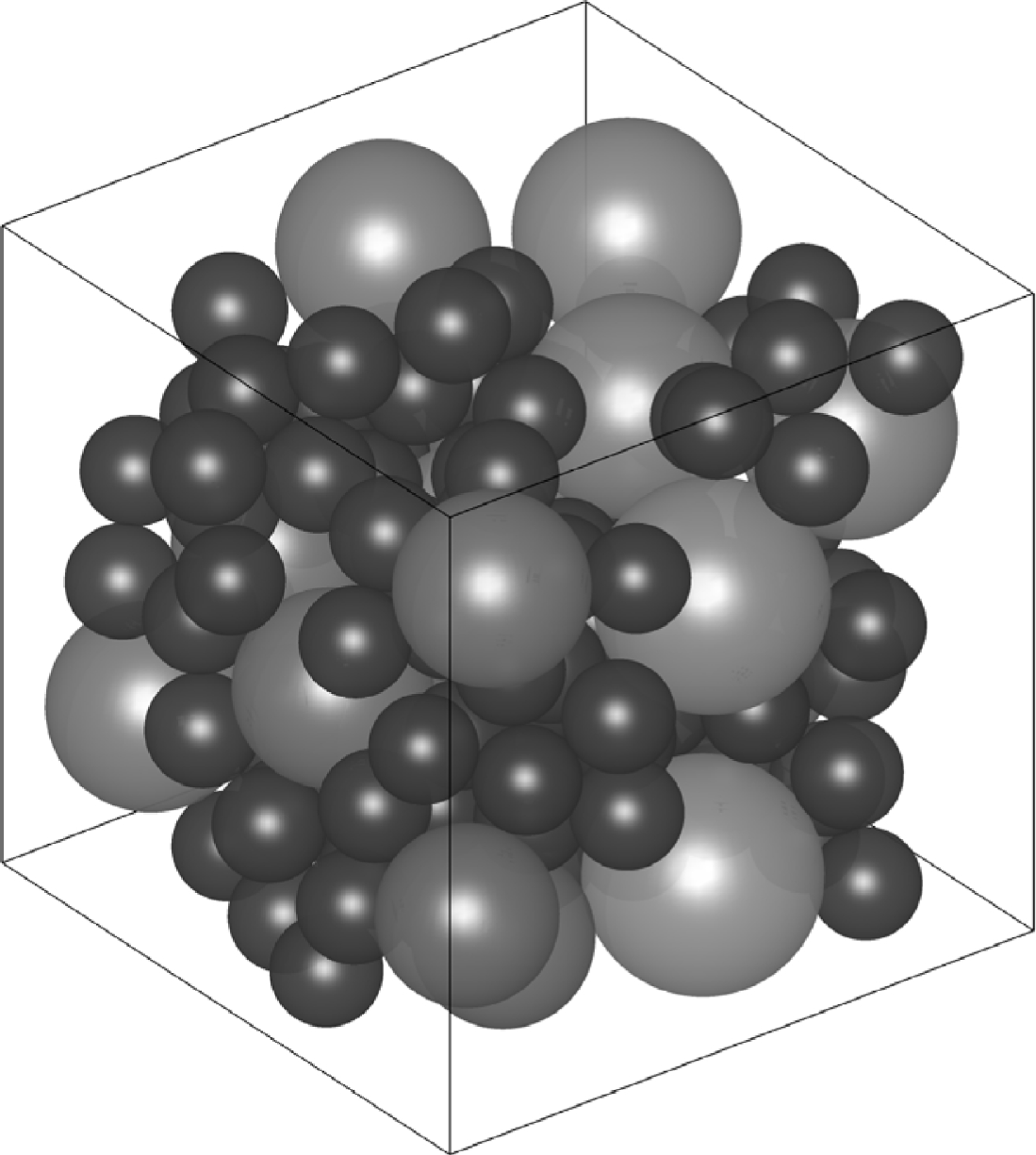}
    &
    \hspace{.30in}
    &
    \includegraphics[scale=0.45]{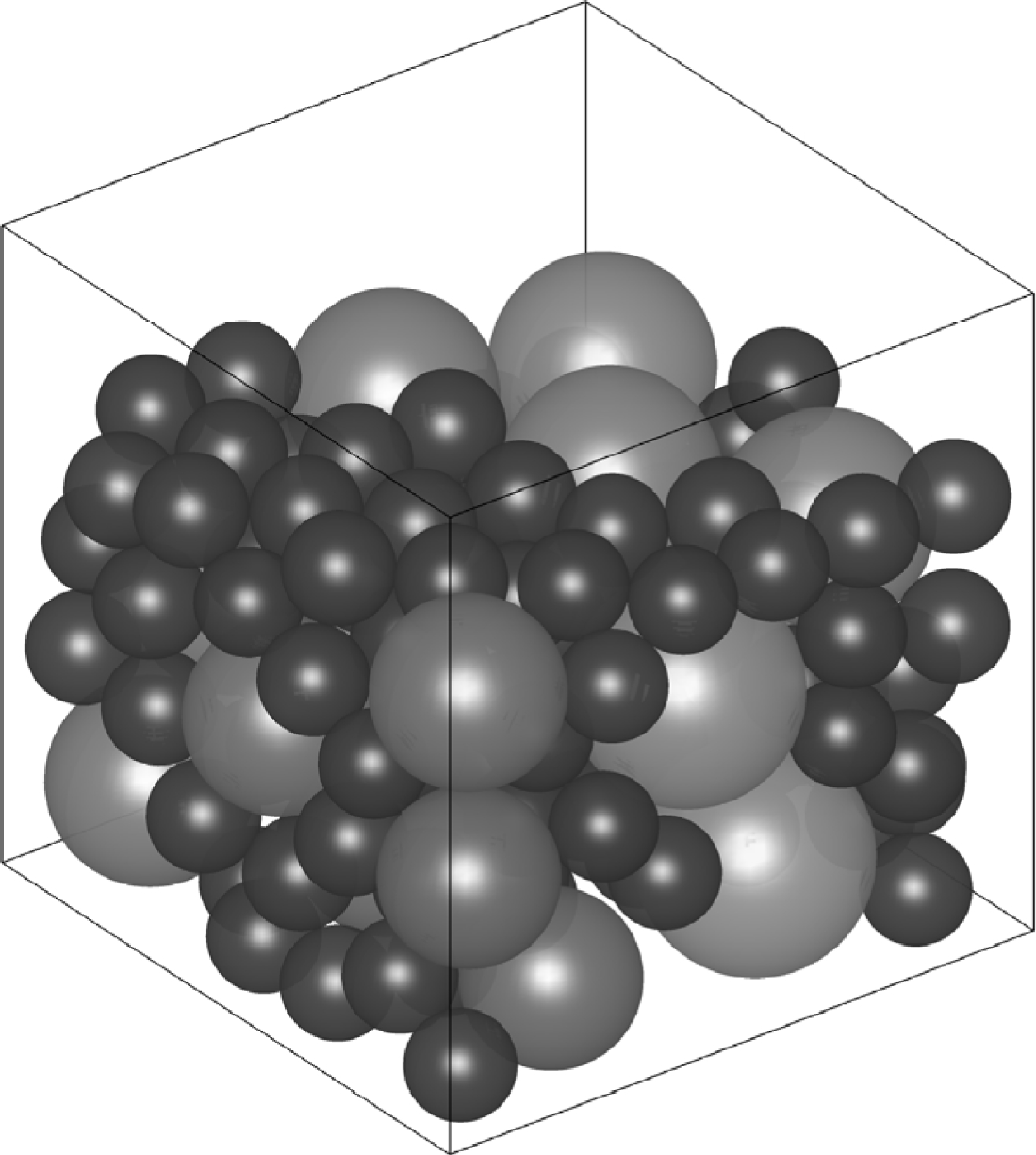}
    \end{tabular}
    \\
    \begin{tabular}{cccc}
    \includegraphics[scale=0.25]{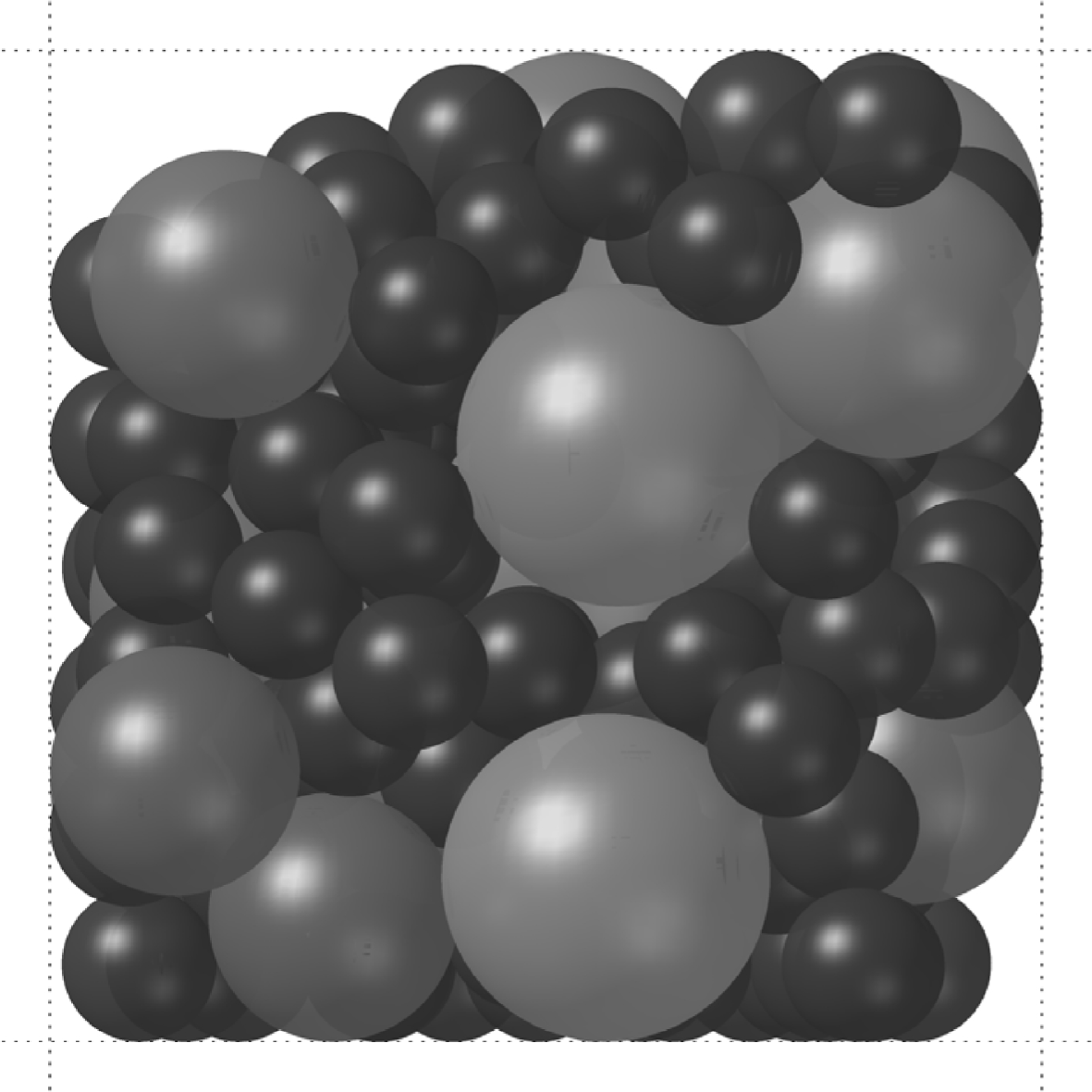}
    &
    \includegraphics[scale=0.25]{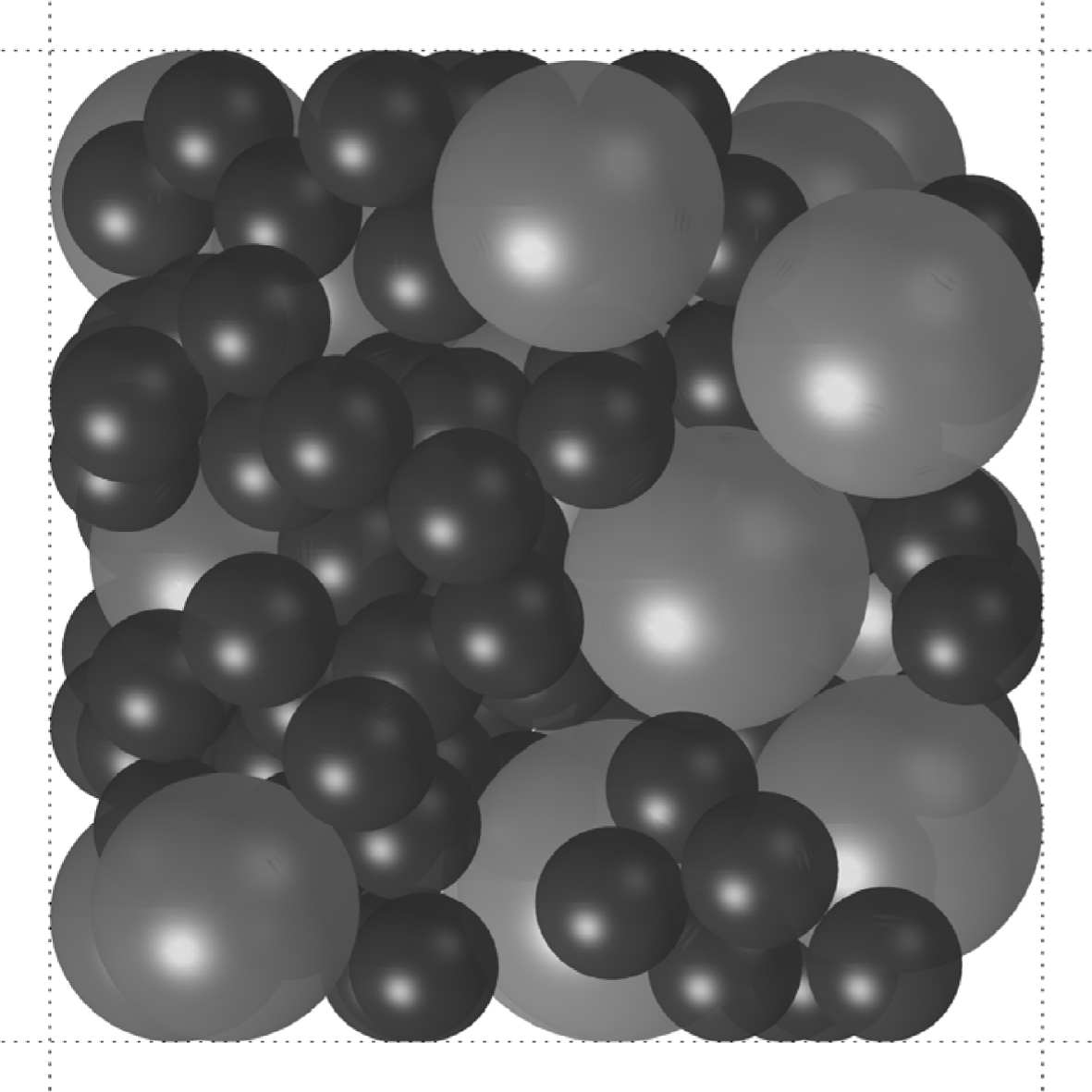}
    &
    \includegraphics[scale=0.25]{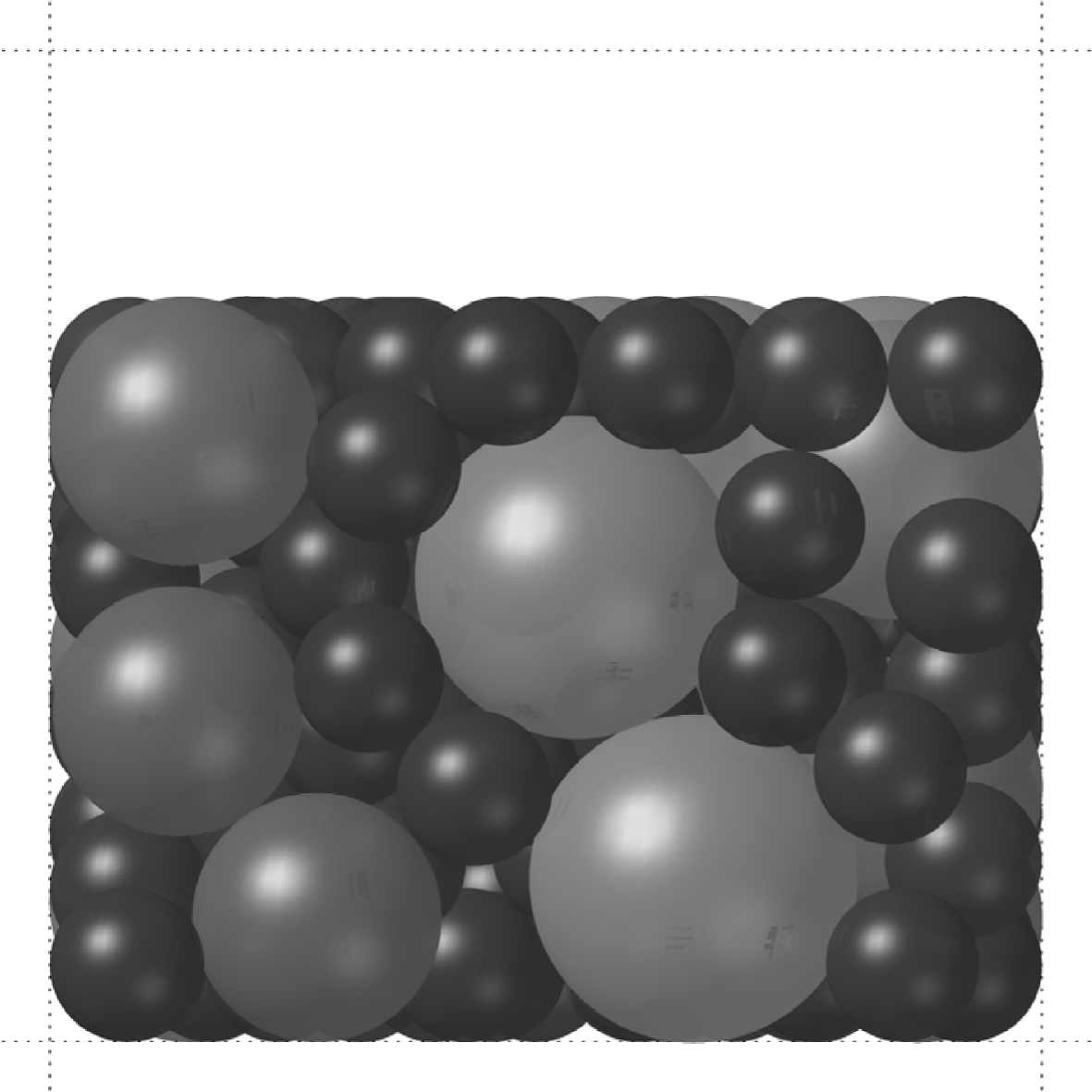}
    &
    \includegraphics[scale=0.25]{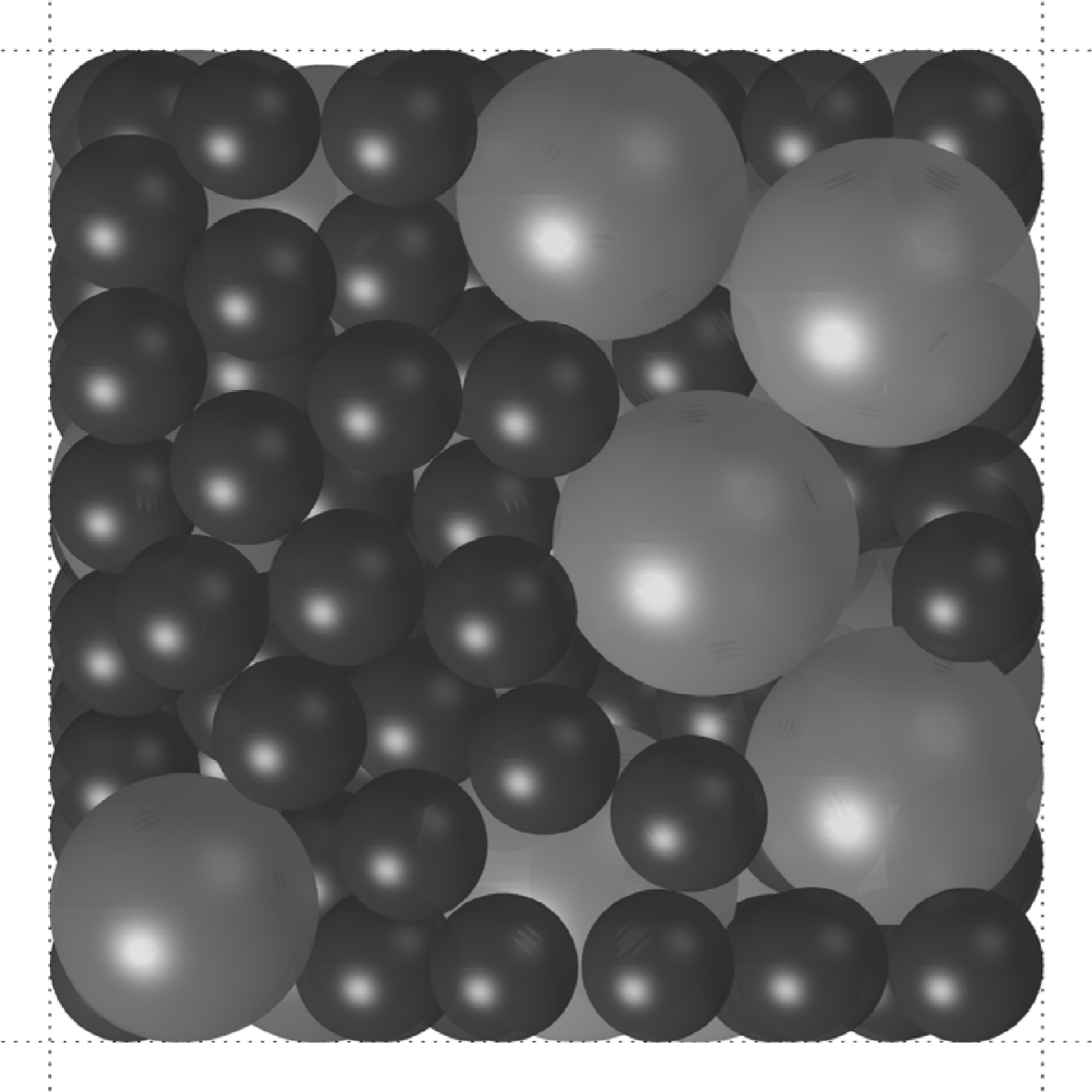}
    \\
    \small{Lateral View}
    &
    \small{Top View}
    &
    \small{Lateral View}
    &
    \small{Top View}
    \end{tabular}
    \caption{Left: Initial granular bed with a relative density of $\rho_{R}=0.42$. Materials are depicted with different shades of grey. Right: Granular bed during consolidation, $\rho_{R}=0.56$. Particle rearrangement is evident.}
    \label{Fig-ConfigGM-Compact}
\end{figure}

The understanding of nonlocal mesoscopic-deformation effects in the macroscopic behavior of disordered granular media is mostly application-driven. In this work, we focus on the consolidation and compaction of confined granular systems (see, e.g., Redanz and Fleck \cite{Redanz-2001}). Specifically, we study the compaction of a powder bed comprised of 125 spherical particles of two different materials. The particle size and elastic properties of the two powders are: i) $E_1=19$~GPa, $\nu_1=0.30$, a log-normal distribution of particle sizes with cut-offs at radiuses $80~\mu$m and $75~\mu$m (typical properties of, e.g., cellulose), ii) $E_2=7$~GPa, $\nu_2=0.20$, a log-normal distribution of particle sizes with cut-offs at radiuses $167~\mu$m and $125~\mu$m (typical properties of, e.g., lactose). The die has a square cross-section of 1~mm~$\times$~1~mm. The initial condition of the un-compacted granular bed is 1~mm in height and it is numerically generated by means of a ballistic deposition technique \cite{Jullien-1989}. Thus, the initial granular bed has a relative density of $\rho_R=0.42$. The left side of Figure~\ref{Fig-ConfigGM-Compact} shows the un-compacted granular bed, materials are depicted with different shades of grey. The right side of the figure shows a snapshot of the powder bed at $\rho_{R}=0.56$ with some particle rearrangement typical of the consolidation process.

\begin{figure}[htbp]
    \centering
    \begin{tabular}{ccc}
    \includegraphics[scale=0.560]{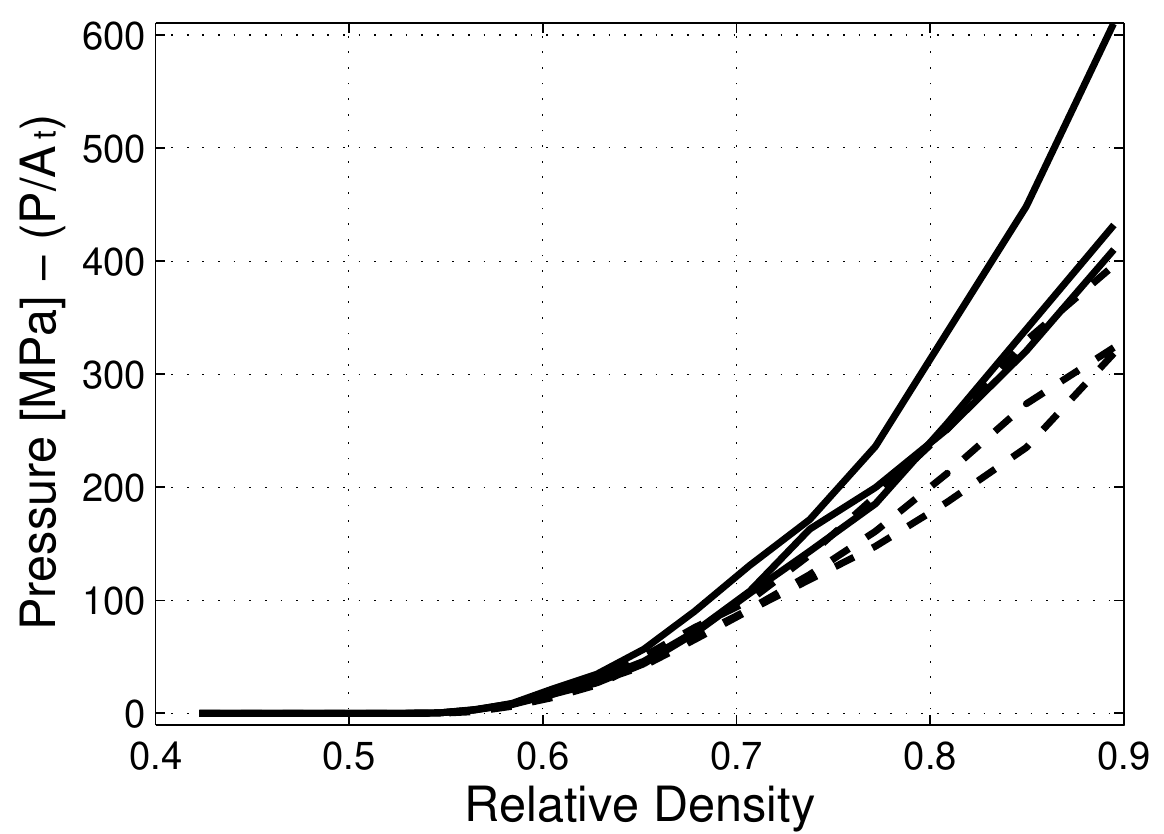}
    &
    \hspace{.20in}
    &
    \includegraphics[scale=0.38]{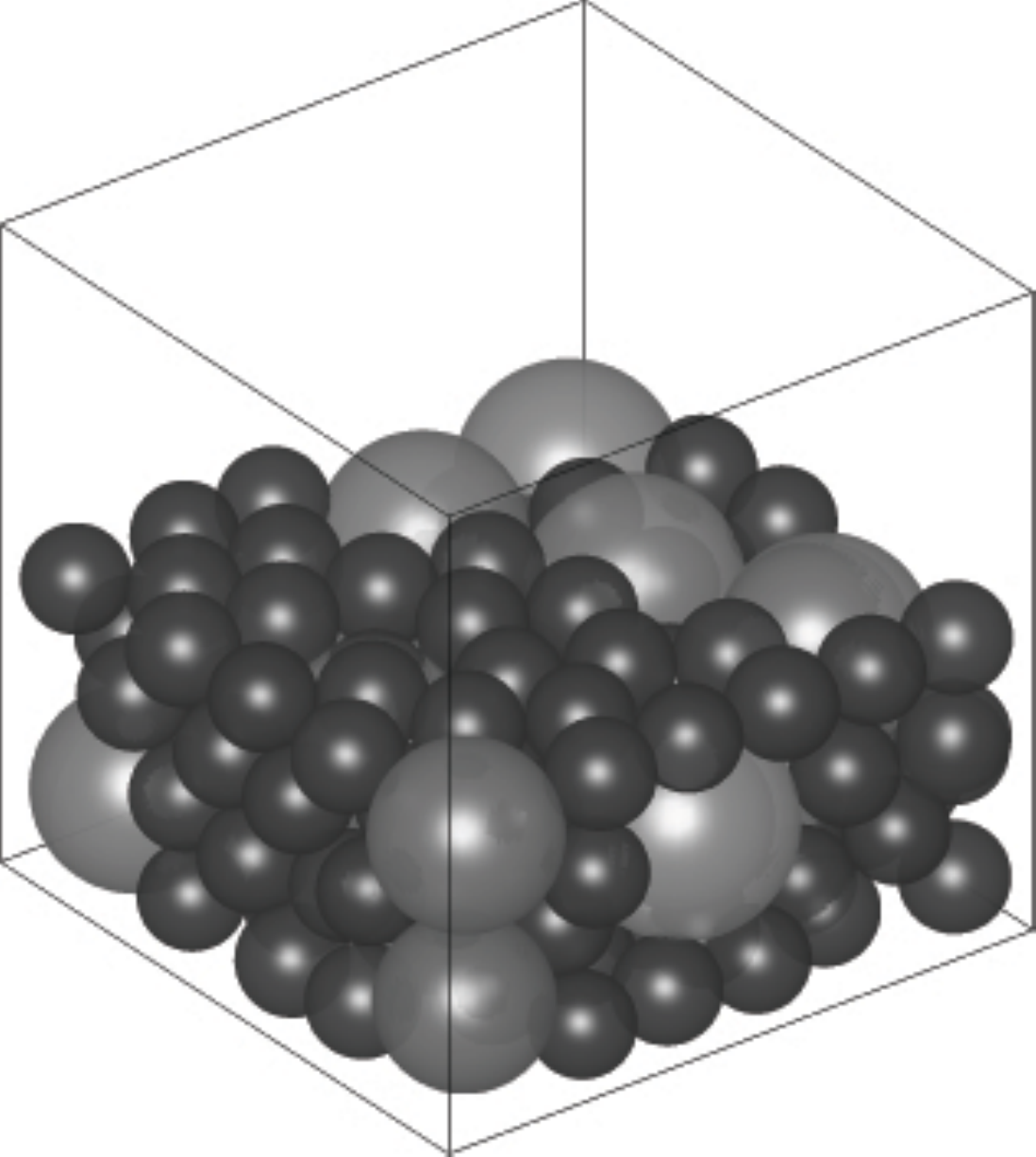}
    \end{tabular}
    \caption{Consolidation and compaction of the disordered heterogeneous granular bed depicted in Figure~\ref{Fig-ConfigGM-Compact}. Left: Applied load, and lateral reaction, versus relative density predicted by Hertz theory (black-dashed curve) and by the nonlocal contact formulation (black curve). Right: Granular bed for $\rho_{R}=0.80$.}
    \label{Fig-StaticGM-ResultsDie}
\end{figure}

The left side of Figure~\ref{Fig-StaticGM-ResultsDie} shows the applied load, and lateral reactions, versus relative density predicted by Hertz theory and by the nonlocal contact formulation. It is interesting to note that both formulations predict a similar deformation pattern during the consolidation process, that is relative density is increased by particle rearrangement as opposed to particle deformation. However, after consolidation is completed, mechanical compaction is required to increase relative density. Therefore, mesoscopic-deformation effects start to play a relevant role in the behavior of the confined granular system and the predictions of Hertz theory depart from those of the nonlocal contact formulation. Again, as it was the case for highly packed granular lattices, the discrepancy between both formulations is significant and evident in the figure. Finally, the granular bed for $\rho_{R}=0.80$ is depicted on the right side of Figure~\ref{Fig-StaticGM-ResultsDie}.

\section{Summary and discussion}
\label{Section-Summary}

We have developed a new nonlocal formulation of contact mechanics that accounts for the interplay of deformations due to multiple contact forces acting on a single particle. This property is achieved by considering the contribution to each contact interface of the nonlocal mesoscopic deformations induced by all other contact forces acting on the particles. As a consequence, the formulation presented here overcomes the unrealistic assumption that contacts are independent regardless the confinement of the granular system---typical of standard contact mechanics theories. For definiteness, we have restricted attention to elastic spheres in the absence of gravitational forces, adhesion or friction. Hence, a notable feature of the nonlocal formulation is that, when nonlocal effects are neglected, it reduces to Hertz theory. We have shown that, up to moderate macroscopic deformations, the predictions of the nonlocal contact formulation are in remarkable agreement with detailed finite-element simulations and experimental observations, and in large disagreement with Hertz theory. The discrepancy between Hertz theory and the extended theory presented in this work has been borne out by studying selected confined granular systems. Specifically, we have investigated different dimensional configurations, different packings and different volumetric confinements of periodic homogeneous systems, and the consolidation and compaction of confined disordered heterogeneous systems. For all studied cases, the discrepancy between both formulations suggests that mesoscopic-deformation effects play a relevant role in the behavior of the granular system. Characteristically, for nearly incompressible materials (i.e., $\nu \rightarrow 1/2$), discrepancies of $40\%$ are observed in periodic bi-dimensional granular configurations under $10\%$ of macroscopic deformations, discrepancies of $200\%$ are predicted for nearly full compaction of body centered cubic granular crystals. Discrepancies in the order of $50\%$ are observed during consolidation and compaction of an heterogeneous and disordered granular bed comprised by relatively compressible materials (i.e., $\nu \sim 1/4$).

We close by pointing out some limitations of our approach and possible avenues for extensions of the formulation.

First, we have restricted attention to linear-elastic isotropic smooth spherical particles. The systematic extension of the nonlocal contact formulation to nonlinear material models, such as neo-Hookean and elasto-plastic constitutive relations, to finite deformations, and to bonding interactions are worthwhile directions of future research.

Second, it is clear that the presented approximations to the single-valued boundary value problem are not the only---perhaps even the best---that are amenable to an efficient closed-form solution. The investigation of other approximation strategies and the determination of the most accurate approximate solution are promising research directions.

Third, although the application of the nonlocal contact formulation to dynamic problems is straightforward, we have only focused on static analyses. The effect of nonlocal mesoscopic deformations in the dynamic macroscopic behavior of confined granular systems remains as an interesting topic to study.

Fourth, it is worth to note the similarity with which discrete granular and atomistic systems are theoretically described despite their different time and length scales. Specifically, iteratomic forces derived from pair potentials in discrete atomistic simulations are the counterparts of the contact forces obtained from traditional contact mechanics theories. In the context of our work, the nonlocal formulation has conceptual similarities with embedded-atom potentials which are composed by pairwise terms and many-body terms \cite{Daw-1984}. Indeed, the many-body terms account for the superposition of electron densities of the surrounding atoms. However, a rigorous analysis of such correlation is desirable, if beyond the scope of this paper.

\section*{Acknowledgements}

The authors gratefully acknowledge the support received from the NSF ERC grant number EEC-0540855, ERC for Structured Organic Particulate Systems.

\section*{Appendix}
\label{Section-Appendix}

We proceed to present two approximate problems to the boundary-value problem depicted in Figure~\ref{Fig-Superposition}c that admit efficient closed-form solutions.

\begin{figure}[htbp]
    \centering
    \begin{tabular}{ccc}
    \includegraphics[scale=0.26]{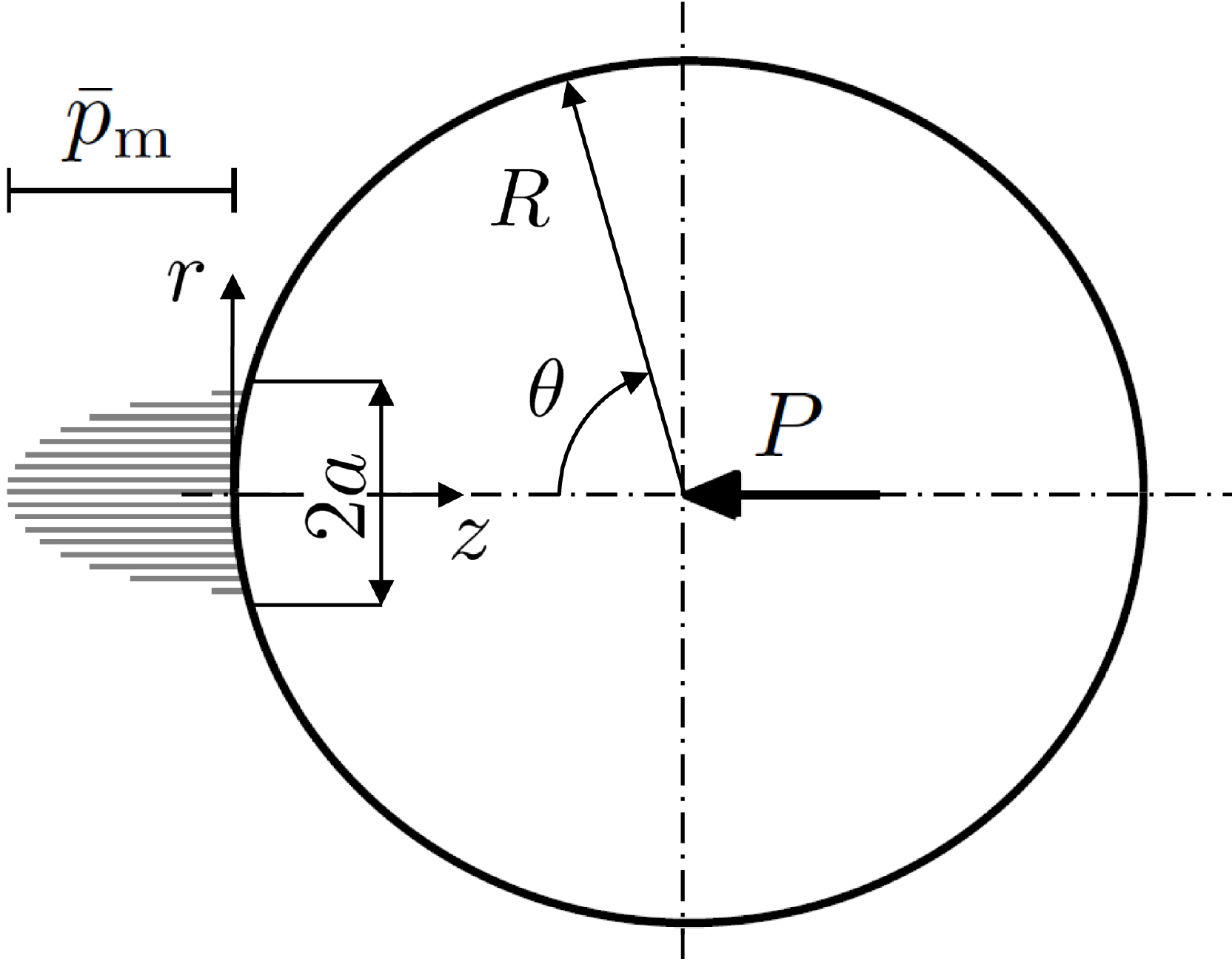}
    &
    \includegraphics[scale=0.26]{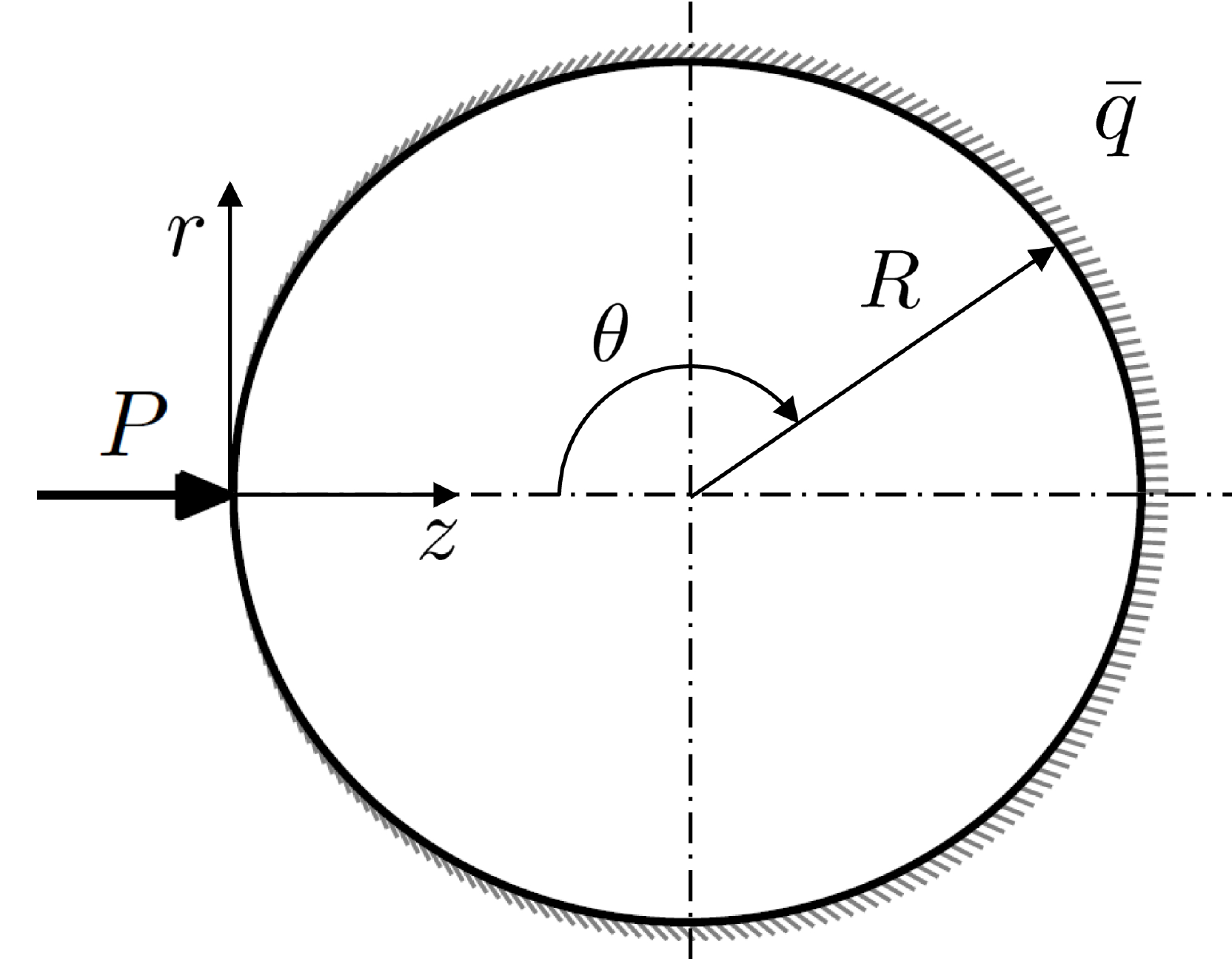}
    &
    \includegraphics[scale=0.26]{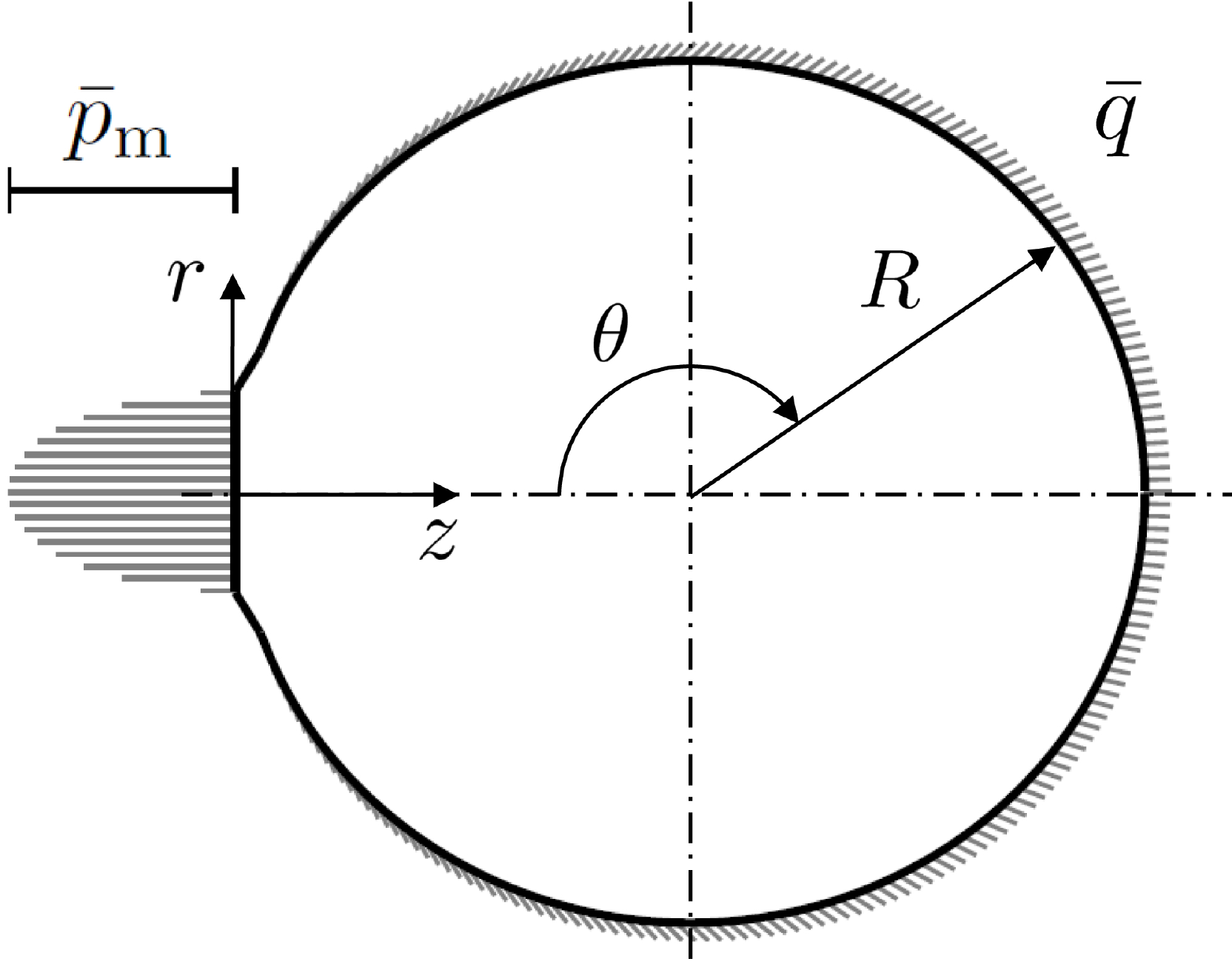}
    \\
    \small{a) Exact boundary-value problem.}
    &
    \small{b) Case I.}
    &
    \small{c) Case II.}
    \end{tabular}
    \caption{Elastic sphere under the action of contact stresses and supported at its center, for $a/R=0.233$ and $\nu=0.50$. Loading configuration of the exact boundary-value problem (a), and the two proposed approximate problems referred to as Case I (b) and Case II (c). The surface traction scales as $\|\bar{q}\|/\bar{p}_\mathrm{m} = \mathcal{O}(a^2/R^2)$.}
    \label{Fig-ApproxSurfaceForces01}
\end{figure}

\subsection*{Case I}

The first approximate problem is the one employed in this paper and corresponds to an elastic spherical particle under a concentrated force $P$---applied at the origin of cylindrical coordinates $(z,r)$ and aligned with the $z$ axis---and supported by a small surface traction $\bar{q}$ given by Equation~\eqref{Eqn-SurfaceTraction}. Figures~\ref{Fig-ApproxSurfaceForces01}a-b illustrate the approximate problem and the exact problem for an incompressible material and $a/R=0.233$, where $a$ is the radius of the contact surface and $R$ is the radius of the elastic sphere.

In order to render the approximate problem exactly solvable, we adopt the displacement and stress fields of a semi-infinite elastic body $\mathcal{B}$ under a concentrated force $P$ applied at the origin of cylindrical coordinates, i.e., the Boussinesq solution \cite{Johnson-1985,Timoshenko-1970}. We then restrict attention to a spherical subbody $\mathcal{B}_{\mathrm{Sph}(R)}$ of radius $R$ and tangent to the $z$-plane at the application point of the force (see Figure~\ref{Fig-ApproxSurfaceForces01}b). The surface points that describe the sphere, i.e., points $(z,r) \in \partial \mathcal{B}_{\mathrm{Sph}(R)}$, are given by $(2R\sin^2(\theta/2),R\sin(\theta))$, and the unit normal to $\partial \mathcal{B}_{\mathrm{Sph}(R)}$ is $n=(-\cos(\theta),\sin(\theta))$. Next, we replace the part of the body outside the sphere, i.e., $\mathcal{B}-\mathcal{B}_{\mathrm{Sph}(R)}$, by the equilibrium surface traction $\bar{q}$ acting on $\partial \mathcal{B}_{\mathrm{Sph}(R)}$, that is
\begin{equation}
    \bar{q}
    =
    -t^{(n)}
    =
    -
    n
    \cdot
    \left(
      \begin{array}{cc}
        \sigma_{zz} & \sigma_{rz} \\
        \sigma_{rz} & \sigma_{rr} \\
      \end{array}
    \right)
\end{equation}
where the stress components correspond to the Boussinesq solution for a semi-infinite elastic body under a concentrated force.

Based on Boussinesq's solution, the displacement of a surface point in $\partial \mathcal{B}_{\mathrm{Sph}(R)}$ is given by
\begin{equation}
    \begin{aligned}
        u_z(\theta) &= \frac{P}{2\pi R E} \left[ \frac{(1+\nu)\sin(\theta/2)}{2} + \frac{1-\nu^2}{\sin(\theta/2)} \right]
        \\
        u_r(\theta) &= \frac{P(1+\nu)}{2\pi R E \sin(\theta)} \left[ \frac{\sin^2(\theta)}{4\sin(\theta/2)} - (1-2\nu)(1-\sin(\theta/2)) \right]
    \end{aligned}
\end{equation}
where $E$ is the Young modulus and $\nu$ is the Poisson's ratio. The components of the stress field---relevant to our analysis---at such surface point are
\begin{equation}
    \begin{aligned}
        \sigma_{zz}(\theta)
        &=
        -\frac{3 P \sin(\theta/2)}{8\pi R^2}
        \\
        \sigma_{rr}(\theta)
        &=
        -\frac{3 P}{8\pi R^2}
        \left[
        \frac{\cos^2(\theta/2)}{\sin(\theta/2)}
        -
        \frac{1-2\nu}{3\sin^2(\theta/2)(1+\sin(\theta/2))}
        \right]
        \\
        \sigma_{rz}(\theta)
        &=
        -\frac{3 P \cos(\theta/2)}{8\pi R^2}
    \end{aligned}
\end{equation}
Thus, after some trite manipulations, the surface traction $\bar{q}$ simplifies to
\begin{equation}
    \bar{q}(\theta)
    =
    \frac{P}{8\pi R^2}
    \left(
        -3 \sin(\theta/2)
        ,
        \frac{1-2\nu}{2\tan(\theta/2)(1+\sin(\theta/2))}-\frac{3\cos(\theta/2)}{2}
    \right)
\end{equation}
and the displacement of a surface point in the normal direction is given by
\begin{equation}
    u_n(\theta)
    =
    (u_z,u_r) \cdot n
    =
    \frac{1+\nu}{4\pi R E} \frac{P \left[ -2(1-\nu)-2(1-2\nu)\sin(\theta/2)+(7-8\nu)\sin^2(\theta/2) \right]}{\sin(\theta/2)}
\end{equation}
Naturally, the above equations correspond to Equations~\eqref{Eqn-SurfaceTraction} and \eqref{Eqn-NonlocalDisplacement} presented in Section~\ref{Section-Formulation}. Finally, it bears emphasis that the surface traction scales as
\begin{equation}
    \frac{\|\bar{q}\|}{\bar{p}_\mathrm{m}} = \mathcal{O}\left(\frac{a^2}{R^2}\right)
\end{equation}
where $\bar{p}_\mathrm{m}$ is the maximum value of the contact pressure (e.g., $\bar{p}_\mathrm{m}=3P/2\pi a^2$ for a spherically-distributed contact pressure). It is also important to note that the displacement reference frame employed in the derivation is the one of Boussinesq's solution (i.e., $u_n(\pi) \rightarrow 0$ as $R \rightarrow \infty$).

\subsection*{Case II}

The second approximate problem aims at a more accurate representation of the contact pressure. We assume a spherically-distributed contact pressure $\bar{p}(r)=\bar{p}_\mathrm{m}\sqrt{1-r^2/a^2}$ applied on a flat circular surface, with radius $a$, of an otherwise spherical elastic particle of radius $R$. Similarly to the previous case, the sphere is supported by a small surface traction $\bar{q}$ that scales as $\|\bar{q}\|/\bar{p}_\mathrm{m} = \mathcal{O}(a^2/R^2)$. Figure~\ref{Fig-ApproxSurfaceForces01}c illustrates the approximate problem for an incompressible material and $a/R=0.233$.

Following the strategy used in Case I, we adopt the displacement and stress field of a semi-infinite elastic body $\mathcal{B}$ under a spherically-distributed pressure applied on a circular surface $\mathcal{S}_\mathrm{c}$ with radius $a$, i.e., Love's solution based on the \emph{potential method} \cite{Love-1929}. We then restrict attention to a spherical subbody $\mathcal{B}_{\mathrm{Sph}(R)}$ of radius $R$, slightly modified to accommodate the flat surface $\mathcal{S}_\mathrm{c}$ (see Figure~\ref{Fig-ApproxSurfaceForces01}c). The surface points of interest to this analysis are give by $(z,r) = (2R\sin^2(\theta/2),R\sin(\theta))$, with unit normal $n=(-\cos(\theta),\sin(\theta))$, for $\theta \in (\arcsin(a/R);\pi]$. Next, we replace the part of the semi-infinite body outside $\mathcal{B}_{\mathrm{Sph}(R)}$ by the equilibrium surface traction $\bar{q}$ acting on $\partial \mathcal{B}_{\mathrm{Sph}(R)} - \mathcal{S}_\mathrm{c}$, and we respect the displacement reference frame employed in Love's solution.

Love's solution for a spherically-ditributed pressure applied on a circular surface with radius $a$ depends on the Newtonian potential $V$ and the Boussinesq's logarithmic potential $\chi$ \cite{Love-1929}. Specifically, the displacements in cylindrical coordinates are
\begin{equation}
    \begin{aligned}
    u_z(z,r)
    &=
    \frac{1+\nu}{2\pi E} \left[ 2(1-\nu) V - z \frac{\partial V}{\partial z} \right]
    \\
    u_r(z,r)
    &=
    -\frac{1+\nu}{2\pi E}\left[ (1-2\nu)\frac{\partial\chi}{\partial r} + z \frac{\partial V}{\partial r} \right]
    \end{aligned}
\end{equation}
and the stress components of interest are
\begin{equation}
    \begin{aligned}
    \sigma_{rr}(z,r)
    &=
    \frac{1}{2\pi}
    \left[
    2\nu\frac{\partial V}{\partial z}
    -
    (1-2\nu) \frac{\partial^2 \chi}{\partial r^2}
    -
    z \frac{\partial^2 V}{\partial r^2}
    \right]
    \\
    \sigma_{zz}(z,r)
    &=
    \frac{1}{2\pi}
    \left[
    \frac{\partial V}{\partial z}
    -
    z \frac{\partial^2 V}{\partial z^2}
    \right]
    \\
    \sigma_{rz}(z,r)
    &=
    -\frac{z}{2\pi}
    \frac{\partial^2 V}{\partial r \partial z}
    \end{aligned}
\end{equation}
In these equations, the Newtonian potential and its derivatives are
\begin{equation}
    \begin{aligned}
    V(z,r)
    &=
    \frac{3P}{4a^2}
    \left[
    \left(2a-\frac{r^2}{a}+\frac{2z^2}{a}\right) \arctan(a/\sqrt{\kappa})
    +
    \frac{r^2 \sqrt{\kappa}}{a^2+\kappa}
    -
    \frac{2z^2}{\sqrt{\kappa}}
    \right]
    \\
    \frac{\partial V}{\partial r}(z,r)
    &=
    -\frac{3Pr}{2a^3}
    \left[
    \arctan(a/\sqrt{\kappa}) - \frac{a\sqrt{\kappa}}{a^2+\kappa}
    \right]
    \\
    \frac{\partial V}{\partial z}(z,r)
    &=
    -\frac{3Pz}{a^3}
    \left[
    \frac{a}{\sqrt{\kappa}}-\arctan(a/\sqrt{\kappa})
    \right]
    \\
    \frac{\partial^2 V}{\partial z^2}(z,r)
    &=
    -\frac{3P}{a^3}
    \left[
    \frac{a}{\sqrt{\kappa}} - \arctan(a/\sqrt{\kappa})
    \right]
    +
    \frac{3 P z^2}{(\kappa^2+a^2 z^2)\sqrt{\kappa}}
    \\
    \frac{\partial^2 V}{\partial r^2}(z,r)
    &=
    -\frac{3P}{2 a^3}
    \left[
    \arctan(a/\sqrt{\kappa}) - \frac{a \sqrt{\kappa}}{a^2+\kappa}
    \right]
    +
    \frac{3 P (\kappa - z^2) \sqrt{\kappa}}{(\kappa^2+a^2 z^2)(a^2+\kappa)}
    \\
    \frac{\partial^2 V}{\partial r \partial z}(z,r)
    &=
    \frac{3 P r z \sqrt{\kappa}}{(\kappa^2+a^2 z^2)(a^2+\kappa)}
    \end{aligned}
\end{equation}
and the derivatives of the Boussinesq's logarithmic potential are
\begin{equation}
    \begin{aligned}
    \frac{1}{r}\frac{\partial\chi}{\partial r}(z,r)
    &=
    \frac{P}{r^2}
    +
    \frac{3Pz}{2a^3}
    \left[
    \frac{a}{\sqrt{\kappa}}-\arctan(a/\sqrt{\kappa})
    \right]
    -
    \frac{Pz}{r^2 \sqrt{\kappa}} \frac{3\kappa-z^2}{2\kappa}
    \\
    \frac{\partial^2 \chi}{\partial r^2}(z,r)
    &=
    -\frac{P}{r^2}
    +
    \frac{3Pz}{2a^3}
    \left[
    \frac{a}{\sqrt{\kappa}}-\arctan(a/\sqrt{\kappa})
    \right]
    +
    \frac{Pz}{r^2 \sqrt{\kappa}} \frac{3\kappa-z^2}{2\kappa}
    \end{aligned}
\end{equation}
with $\kappa$ given by the positive root of $\frac{r^2}{a^2+\kappa}+\frac{z^2}{\kappa}=1$.

Since the analytical evaluation of the surface traction $\bar{q}(\theta)$ and the normal component of the surface displacement $u_n(\theta)$ does not result in any further simplification of the equations presented above, we will only proceed to compare these results with those of the exact solution by means of numerical experimentation. We present this comparative study next in turn.

\subsection*{Accuracy assessment}

The accuracy of the two approximate problems presented in this Appendix is assessed by comparing their predictions with the exact solution recently presented by Zhupanska \cite{Zhupanska-2011}. Specifically, an elastic incompressible sphere that accommodates a flat circular area of $a/R=0.233$ is employed. Since it is assumed that $a \ll R$, this example corresponds to a non-small contact area and therefore it allows for exploring the limits of validity of the presented formulation. Likewise, the selection of an incompressible material corresponds to the scenario with the largest displacements.

The exact solution reported by Zhupanska \cite{Zhupanska-2011} uses a general solution for the axisymmetric problem of an elastic sphere and a dual series equation approach to reduce the original boundary-value problem to a Fredholm integral equation of the second kind, which is then solved numerically. The analytical solution is achieved without replacing the contact area of the sphere by an elastic half-space and it allows for determining the contact pressure and the displacement of the traction-free surface. The results also show that the spherically-distributed contact pressure predicted by Hertz theory remains accurate for relatively large contact areas (i.e., the relative error in $\bar{p})_\mathrm{m}$ is smaller than $6\%$ for $a/R < 0.40$, provided linear elasticity remains valid).

\begin{figure}[h]
    \centering
    \includegraphics[scale=0.56]{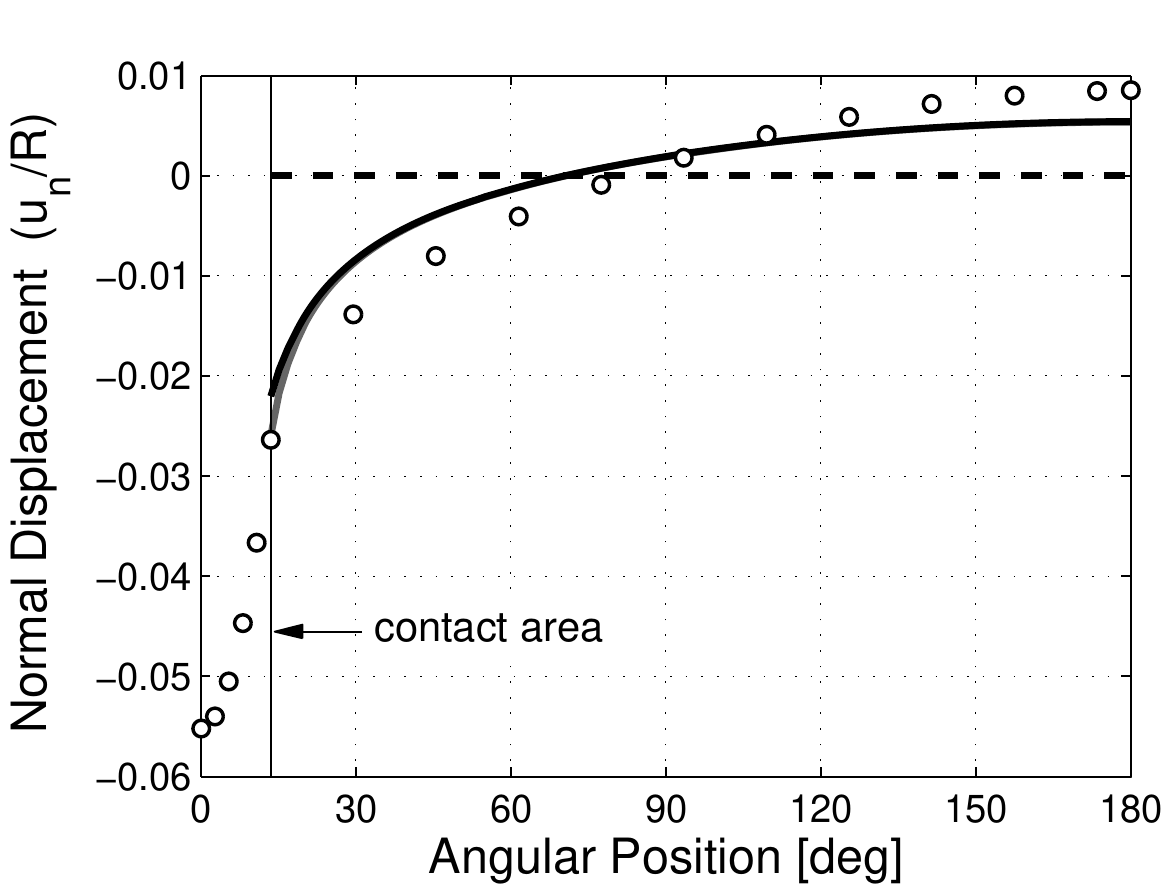}
    \caption{Normal displacement of surface points for an elastic sphere under the action of contact stresses and supported at its center, for $a/R=0.233$ and $\nu=0.50$. The exact solution (symbols) is taken from Zhupanska \cite{Zhupanska-2011}. The approximate solutions correspond to approximate problems referred to as Case I (black solid curve) and Case II (grey solid curve). The dashed curve corresponds to the assumption of traditional contact mechanics theories.}
    \label{Fig-ApproxProblem-Accuracy}
\end{figure}

Figure~\ref{Fig-ApproxProblem-Accuracy} shows the exact solution and the approximate solutions that correspond to the approximate problems referred to as Case I and Case II. The results are is very good agreement and also reveal Case I as being equally effective, and more efficient, than Case II.

\bibliographystyle{plainnat}

\end{document}